\documentclass[aps,pra,twocolumn,showpacs,superscriptaddress]{revtex4}
\usepackage{graphicx}
\usepackage{amsmath}
\usepackage{amssymb}

\input{epsf}
\begin{document}

\title{Energy spectrum of
harmonically trapped two-component Fermi gases:
Three- and Four-Particle Problem}

\author{K.~M. Daily}
\affiliation{Department of Physics and Astronomy,
Washington State University,
  Pullman, Washington 99164-2814, USA}
\author{D. Blume}
\affiliation{Department of Physics and Astronomy,
Washington State University,
  Pullman, Washington 99164-2814, USA}

\date{\today}

\begin{abstract}
Trapped two-component Fermi gases allow for the investigation
of the so-called BCS-BEC crossover by tuning the interspecies 
atom-atom $s$-wave 
scattering length scattering $a^{(aa)}$ from attractive to
repulsive, including vanishing and infinitely large values.
Here, we numerically determine 
the energy spectrum
of the equal-mass spin-balanced four-fermion system---the
smallest few-particle system that exhibits
BCS-BEC crossover-like behavior---as 
a function of $a^{(aa)}$ 
using the stochastic variational approach. 
For comparative purposes, we also treat the two- and three-particle systems.
States with vanishing and finite total angular momentum 
as well as with natural and unnatural parity are considered.
In addition,
the energy spectrum of weakly-attractive
and weakly-repulsive 
gases is 
characterized by employing a perturbative framework that 
utilizes hyperspherical coordinates. 
The hyperspherical
coordinate approach allows 
for the straightforward assignment of 
quantum numbers and furthermore provides 
great insights into the strongly-interacting unitary regime.
\end{abstract}


\maketitle

\section{Introduction}
\label{sec_intro}
Few-body systems show rich behaviors that range from the realization
of
highly-correlated states to weakly-bound Borromean states,
and have long been of great interest to 
chemists as well as nuclear and atomic physicists.
To date,
the determination of the entire energy spectrum,
or parts thereof, of small bosonic or fermionic
systems consisting of four or more
constituents remains a challenge despite the ever increasing 
computational resources.
Recently, significant progress has been made in the theoretical
characterization
of weakly-bound bosonic 
tretramers~\cite{plat04,yama06,hann06,hamm07,stec09a,wang09}. 
In particular, 
for each Efimov trimer
there exist two 
tetramer states~\cite{hamm07,stec09a}, which dissociate into
four free bosons at critical negative scattering lengths.
The ratio of the scattering length at which the trimer state
becomes unbound and that at which the first or second tetramer 
state becomes unbound has been predicted to be 
universal~\cite{stec09a}.
Recently, this prediction has been confirmed by 
loss rate measurements 
on the negative scattering length side~\cite{ferl09}. Working
with an atomic Cs sample at temperatures just above the transition
temperature to quantum degeneracy, the Innsbruck group~\cite{ferl09} was 
able to observe enhanced losses at magnetic field strengths
that correspond quite well to the
theoretically predicted scattering length ratios~\cite{stec09a}. 
By now, several groups have reported experimental evidence
for universal four-boson physics~\cite{ferl09,zacc09,poll09}.

While the universal
properties of few-boson systems interacting through short-range potentials
depend
on two atomic physics parameters, i.e., the
two-body $s$-wave scattering length $a^{(aa)}$ and a three-body 
parameter (see, e.g., Ref.~\cite{braa06}), 
the universal
properties of dilute equal-mass two-component Fermi gases
interacting through short-range potentials
with interspecies $s$-wave interactions 
depend only on the $s$-wave scattering 
length~\cite{bake99,ohar02,ho04,tan04,chan05,astr04c,chan04a,thom05,cast04,wern06,son06,stew06,tan08a,tan08b,tan08c,braa08,braa08a,wern10}.
%(see, e.g., Ref.~\cite{gior08}).
%Correspondingly,
%if the range $r_0$ of the underlying two-body potential
%is significantly smaller than 
%the 
%$s$-wave scattering length, then
%the low-energy spectrum of two-component Fermi
%gases
%is expected to be universal, i.e., fully determined by $a^{(aa)}$
%and independent of the details of the underlying two-body 
%potential
Experimentally, small two-component Fermi gases can be realized by loading
a deep three-dimensional
optical lattice with a deterministic number of atoms per 
lattice site~\cite{grei02,koeh05,thal06}.
If the tunneling between neighboring sites is negligible, each lattice 
site can be treated as an independent, approximately harmonically
confined few-fermion system.

This paper determines and characterizes
the energy spectrum of three- and four-particle
equal-mass two-component Fermi gases
as a function of the $s$-wave scattering length
under spherically symmetric
harmonic confinement. 
For this confining geometry, the total angular momentum
$L_{tot}$ and the total parity $\Pi_{tot}$ 
are good quantum numbers
throughout the entire 
BCS-BEC
crossover.
The three-fermion spectra have been discussed 
previously~\cite{wern06a,kest07,stet07,stec08} 
and are
included here primarily for illustrative and comparative purposes.
Our calculations follow two distinctly different avenues. On the one
hand, we numerically 
determine the energy spectrum of few-fermion systems
throughout the entire 
crossover. For the four-fermion system, we employ the stochastic variational
approach~\cite{varg95,varg01,cgbook,sore05,stec07c,stec07b,stec08}.
In contrast to previous 
studies~\cite{stec07c,stec07b,blum07,stec08,blum09a}, we utilize
basis functions with well-defined angular momentum and parity
and determine the eigenenergies for a range of angular momenta.
On the other hand, we determine the eigenspectrum 
semi-analytically within first
order degenerate perturbation
theory. While necessarily limited to small $|a^{(aa)}|$,
this approach allows
for the classification of a large portion of the energy spectrum in
terms of appropriate quantum numbers.
To characterize the energy spectrum in the weakly-interacting regime, we
employ hyperspherical 
coordinates~\cite{aver89,lin95,bohn98,timo02,timo04,ripe05,wern06,ritt06,reviewgreen} 
and write
the non-interacting wave functions
in the relative coordinates
as a product of a hyperangular channel function
and a hyperradial weight function.
The eigenenergies of the non-interacting system have, in general,
large degeneracies, which are
partially lifted by the two-body interactions. The energy splittings
can,
to leading order,
be calculated perturbatively.
Compared to calculations that utilize Cartesian single particle
coordinates, 
one distinct advantage of the hyperspherical approach is 
that certain features carry over, with some modifications, to
the strongly-interacting unitary 
regime~\cite{wern06,blum07,stec08,ritt06,ritt08}.
Our numerically determined spectra 
at unitarity
can thus be interpreted within the hyperspherical framework.

The remainder of this manuscript is organized as follows.
Section~\ref{sec_hamiltonian}
introduces the system Hamiltonian under study
and provides other background information.
Section~\ref{sec_hyperspherical}
discusses the hyperspherical framework
and its implications for the non-interacting, weakly-interacting
and strongly-interacting three- and four-fermion systems.
The numerical basis set type expansion
approaches for the three- and four-fermion
problems are discussed in
Sec.~\ref{sec_threebody} and
Sec.~\ref{sec_stochastic}, respectively.
Section~\ref{sec_results} 
summarizes our numerical and semi-analytical
perturbative 
results for the three- and four-fermion systems.
We discuss the degeneracies and
quantum numbers of the energy levels throughout the BCS-BEC
crossover.
Furthermore, we characterize the energy spectrum at unitarity
and present a simple model that predicts the 
energy spectrum of the three-fermion system and a subset of the 
energy spectrum of the four-fermion system
at unitarity.
Lastly, Sec.~\ref{sec_conclusion} summarizes our main results.

\section{Theoretical background}
\label{sec_theory}
This section introduces the system Hamiltonian and 
discusses our
approaches to determining the eigenspectrum 
of equal-mass two-component Fermi 
gases perturbatively and numerically.
\subsection{System Hamiltonian and other background information}
\label{sec_hamiltonian}
We consider
small equal-mass two-component Fermi gases 
under
external harmonic confinement
consisting of $N$
atoms with mass $m$ 
and position vectors $\vec{r}_j$,
measured with respect to the center of the trap.
Our model Hamiltonian $H$ 
reads
\begin{eqnarray}
\label{eq_ham}
H = 
H^{\mathrm{ni}}
+
V_{\mathrm{int}}(\vec{r}_1,\cdots,\vec{r}_N),
\end{eqnarray}
where the non-interacting Hamiltonian $H^{\mathrm{ni}}$ is given by
\begin{eqnarray}
\label{eq_hamni}
H^{\mathrm{ni}}=
\sum_{j=1}^{N} \left[ -\frac{\hbar^2}{2m} \nabla_{\vec{r}_j}^2 
+ V_{\mathrm{trap}}(\vec{r}_j)
\right] 
\end{eqnarray}
and
the external spherically symmetric harmonic confining potential
$V_{\mathrm{trap}}$ is characterized by the angular trapping frequency $\omega$,
\begin{eqnarray}
\label{eq_trap}
V_{\mathrm{trap}}(\vec{r}_j) = 
\frac{1}{2} m \omega^2 \vec{r}_j^2  .
\end{eqnarray} 
The potential $V_{\mathrm{int}}$ accounts for the
short-range two-body interactions $V_{\mathrm{tb}}$ between unlike atoms,
\begin{eqnarray}
\label{eq_int}
V_{\mathrm{int}}(\vec{r}_1,\cdots,\vec{r}_N)=
\sum_{j=1}^{N_{\uparrow}}  
\sum_{k=N_{\uparrow}+1}^{N} V_{\mathrm{tb}}(\vec{r}_j-\vec{r}_k),
\end{eqnarray}
where the number $N_{\uparrow}$ of spin-up atoms and the number 
$N_{\downarrow}$ of spin-down atoms add up to the total number
of atoms, i.e., $N_{\uparrow}+N_{\downarrow}=N$.
For spin-imbalanced systems, $N_{\uparrow}$ denotes the number of atoms
of the majority species and $N_{\downarrow}$ that of the minority species.
Throughout, we assume that  
the two-body potential $V_{\mathrm{tb}}$ is characterized by the 
$s$-wave atom-atom scattering length $a^{(aa)}$ and possibly a range 
parameter $r_0$. The different functional forms of
$V_{\mathrm{tb}}$ employed in our
calculations are discussed below.
The goal of this paper is to determine and interpret
the eigenenergies $E(N_{\uparrow},N_{\downarrow})$
of the Hamiltonian $H$, Eq.~(\ref{eq_ham}).

If the atom-atom scattering length $a^{(aa)}$ is negative and small in absolute
value, i.e., $|a^{(aa)}| \ll a_{\mathrm{ho}}$, where 
$a_{\mathrm{ho}}$ denotes the oscillator length associated with the atom mass $m$,
\begin{eqnarray}
a_{\mathrm{ho}}=\sqrt{\frac{\hbar}{m \omega}},
\end{eqnarray}
then the Fermi system behaves like a weakly-attractive atomic gas.
In this case, the energy shifts due to
the interactions can be described, to leading order,
within first order degenerate perturbation theory
that treats
$H^{\mathrm{ni}}$ as the unperturbed Hamiltonian
and 
$V_{\mathrm{int}}$ as the perturbation~\cite{stec07b,stec08}.
It is then convenient to 
parametrize the two-body potential $V_{\mathrm{tb}}(\vec{r}_{jk})$
by
Fermi's pseudo-potential $V_{\mathrm{F}}(\vec{r}_{jk})$~\cite{ferm34},
\begin{eqnarray}
\label{eq_fermi}
V_{\mathrm{F}}(\vec{r}_{jk}) = 
\frac{4 \pi \hbar^2}{m} a^{(aa)} \delta(\vec{r}_{jk}),
\end{eqnarray}
which
allows for an analytical 
evaluation of the matrix elements and,
if employed within first order perturbation theory,
does not lead to divergencies.
In
general, multiple eigenfunctions $\psi_j^{\mathrm{ni}}$ of the non-interacting
atomic system are degenerate
so that the first order energy shifts $E^{(1)}$
are obtained by solving the determinantal equation
\begin{eqnarray}
\mbox{det} \left( \underline{V}_{\mathrm{int}} - E^{(1)} \underline{I} \right) =0,
\end{eqnarray}
where $\underline{I}$ denotes the identity matrix
and the matrix elements $(\underline{V}_{\mathrm{int}})_{jk}$
are given by
\begin{eqnarray}
\left( \underline{V}_{\mathrm{int}} \right)_{jk}=
\langle \psi_j^{\mathrm{ni}} | V_{\mathrm{int}} | \psi_k^{\mathrm{ni}} \rangle.
\end{eqnarray}
Here, $j$ and $k$ run from $1$ to $g_{\mathrm{ni}}$, where $g_{\mathrm{ni}}$ denotes
the degeneracy of the eigenenergy $E_{\mathrm{ni}}(N_{\uparrow},N_{\downarrow})$ 
under
consideration.

The possibly most direct approach for constructing
the eigenfunctions $\psi_j^{\mathrm{ni}}$ is
to write the $\psi_j^{\mathrm{ni}}$ as a product of two determinants,
one for the spin-up atoms and one for the spin-down atoms.
The determinants themselves are constructed from the
single-particle wave functions $\phi_{n_p,l_p,m_p}^{\mathrm{SP}}(\vec{r}_p)$,
$p=1, \cdots, N_{\uparrow}$ or $p=N_{\uparrow}+1, \cdots, N$,
which are eigenfunctions of the single-particle
harmonic oscillator Hamiltonian $H^{\mathrm{SP}}_p$,
\begin{eqnarray}
H^{\mathrm{SP}}_p=-\frac{\hbar^2}{2m}\nabla^2_{\vec{r}_p}+V_{\mathrm{trap}}(\vec{r}_p).
\end{eqnarray}
Above, $n_p$, $l_p$ and $m_p$ denote the single particle
radial, orbital angular momentum and projection
quantum numbers, respectively.
For a given 
energy $E_{\mathrm{ni}}(N_{\uparrow},N_{\downarrow})$ 
of the non-interacting many-body system,
the 
single-particle wave functions $\phi_{n_p,l_p,m_p}^{\mathrm{SP}}(\vec{r}_p)$
have to be chosen such that
their eigenenergies
obey the constraint
\begin{eqnarray}
E_{\mathrm{ni}}(N_{\uparrow},N_{\downarrow}) = 
\sum_{p=1}^{N_{\uparrow}} 
\left( 2 n_{p} + l_{p}+\frac{3}{2} \right)
\hbar \omega
 + \nonumber \\
\sum_{p=N_{\uparrow}+1}^{N} 
\left( 2 n_{p} + l_{p}+\frac{3}{2} \right)
\hbar \omega,
\end{eqnarray}
with the additional restriction that the sets of quantum numbers
$(n_p,l_p,m_p)$ and
$(n_q,l_q,m_q)$
differ by at least one entry for $p \ne q$, where
$p,q =1,\cdots,N_{\uparrow}$
or
$p,q =N_{\uparrow}+1,\cdots,N$.
Following this approach, the first order
energy shifts $E^{(1)}$ of the energetically lowest
lying gas-like state for systems
with $N_{\uparrow}-N_{\downarrow}=0,\pm1$ and up to $N=20$ atoms
have been calculated~\cite{stec08}. 

Although in principle straightforward, the outlined construction of
the wave functions of the non-interacting Fermi gas and their
use in evaluating the energy shifts $E^{(1)}$ has several
disadvantages.
The number of degenerate states of the non-interacting system
increases rapidly with increasing energy. For example,
for the three-particle system with $(N_{\uparrow},N_{\downarrow})=(2,1)$,
the lowest four energies 
$E_{\mathrm{ni}}(2,1)=11 \hbar \omega/2$, $13 \hbar \omega/2$,
$15 \hbar\omega/2$ and $17\hbar\omega/2$ 
of the non-interacting system 
have degeneracies $g_{\mathrm{ni}}=3$, $18$, $73$ and $228$,
and the determination of the
energy shifts thus requires the construction 
and diagonalization of increasingly
large interaction potential matrices
$\underline{V}_{\mathrm{int}}$. Furthermore, the outlined construction
includes center-of-mass excitations and does not take
advantage of the fact that the total angular momentum
$L_{tot}$, the corresponding
$z$-projection  
and the parity $\Pi_{tot}$ of the system
are good quantum numbers. Lastly, the anti-symmetrization
is accomplished through the use of determinants, leading to
$N_{\uparrow}! \times N_{\downarrow}!$ 
terms for each $\psi_j^{\mathrm{ni}}$.

This paper
pursues an alternative
approach and writes the non-interacting
wave functions in terms of hyperspherical 
coordinates~\cite{aver89,lin95,bohn98,timo02,timo04,ripe05,wern06,ritt06,reviewgreen}. 
This approach separates off the center-of-mass degrees of freedom,
treats one angular momentum at a time and ensures
the proper anti-symmetry of the wavefunction by utilizing
angular momentum algebra.
Using the wave functions of the non-interacting atomic Fermi gas,
written in terms of hyperspherical coordinates,
we are able to determine the first-order energy shifts for a large
portion of the spectrum of weakly-interacting
equal-mass
two-component atomic 
Fermi gases with $N=3$ and 4 semi-analytically
(see Sec.~\ref{sec_hyperspherical} and Sec.~\ref{sec_results}).

\begin{figure}
\vspace*{+.4cm}
\includegraphics[angle=0,width=70mm]{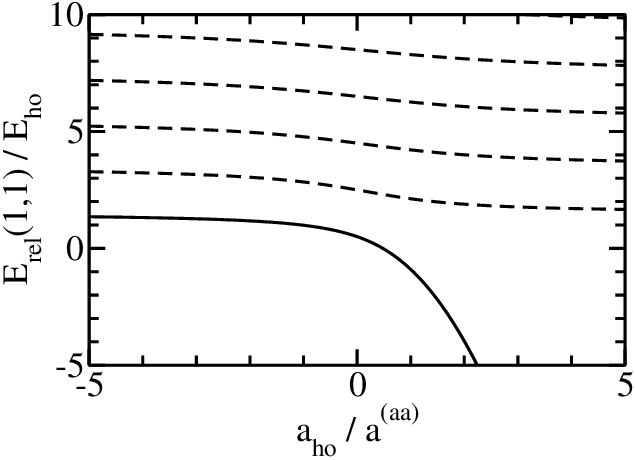}
\vspace*{0.1cm}
\caption{Relative 
$s$-wave energies $E_{\mathrm{rel}}(1,1)$ of the trapped atom-atom 
system, obtained by 
solving Eq.~(\ref{eq_busch})
for $k=aa$, as a function of $a_{\mathrm{ho}}/a^{(aa)}$.
The lowest $s$-wave eigenenergy (solid line)
is referred to as $E_{\mathrm{dimer}}$ in the text.
Here, $E_{\mathrm{ho}}$ denotes the harmonic oscillator
energy, $E_{\mathrm{ho}}=\hbar \omega$.
}\label{fig_energyn2}
\end{figure}

When $a^{(aa)}$ is positive and small
($a^{(aa)} \ll a_{\mathrm{ho}}$), diatomic
bosonic molecules can form and, if this happens,
the 
Fermi system behaves like a weakly-repulsive molecular Bose gas.
In this limit, the dimers or diatomic
molecules can to a good approximation
be treated as bosonic point particles with mass 
$2m$~\cite{astr04c,petr04aa,petr05,stec07b,stec08}
and internal energy $E_{\mathrm{dimer}}$; as detailed below,
this internal energy accounts for the presence of
the external confinement.
The effective model systems for $N=3$ and
$N=4$ then consist of two particles, an atom and a dimer
in the three-particle case
and two dimers
in the four-particle case~\cite{petr03,petr04aa,petr05,stec07b,stec08}. 
Separating off the center-of-mass motion, the dynamics are
governed by the relative
effective 
Hamiltonian $H^{\mathrm{eff}}$,
\begin{eqnarray}
H^{\mathrm{eff}}= -\frac{\hbar^2}{2 \mu^{(k)}} \nabla^2_{\vec{r}} + 
\frac{1}{2} \mu^{(k)} 
\omega^2
\vec{r}^2 + V_{\mathrm{F,reg}}^{(k)}(\vec{r}),
\end{eqnarray}
where $k$ stands for $ad$ (atom-dimer)
and $dd$ (dimer-dimer) for the three- and four-fermion
systems, respectively,
and the position vector $\vec{r}$ denotes the 
atom-dimer and dimer-dimer distance vector
for the three- and four-fermion
systems, respectively.
The reduced masses $\mu^{(ad)}$ and $\mu^{(dd)}$
of the atom-dimer and dimer-dimer systems are defined
as $\mu^{(ad)}=2m/3$ and $\mu^{(dd)}=m$.
In this model,
the atom-molecule and molecule-molecule
interactions are conveniently described
through Fermi's regularized
pseudo-potential $V_{\mathrm{F,reg}}^{(k)}(\vec{r})$~\cite{huan57},
\begin{eqnarray}
\label{eq_fermireg}
V_{\mathrm{F,reg}}^{(k)}(\vec{r}) = \frac{2 \pi \hbar^2}{\mu^{(k)}} 
a^{(k)} \delta(\vec{r}) 
\frac{\partial}{\partial r} r,
\end{eqnarray}
with effective atom-dimer
and dimer-dimer scattering lengths
$a^{(ad)}$ and 
$a^{(dd)}$, 
respectively~\cite{skor57,petr03,mora04,mora05a,stec08,tan08a},
\begin{eqnarray}
\label{eq_atomdimer}
a^{(ad)} \approx 1.1790662349a^{(aa)}
\end{eqnarray}
and~\cite{petr04aa,stec07b,stec08} 
\begin{eqnarray}
\label{eq_dimerdimer}
a^{(dd)} \approx 0.608a^{(aa)}.
\end{eqnarray}
The $s$-wave ($l_{\mathrm{eff}}=0$) 
eigenenergies $E_{\mathrm{eff}}$ of $H^{\mathrm{eff}}$ 
are readily obtained by solving the transcendental equation~\cite{busc98}
\begin{eqnarray}
\label{eq_busch}
\frac{a^{(k)}}{a_{\mathrm{ho},\mu}^{(k)}} = \frac{\Gamma \left(-\frac{E_{\mathrm{eff}}}{2\hbar \omega}+
\frac{1}{4} \right)}
{2\Gamma \left( -\frac{E_{\mathrm{eff}}}{2 \hbar \omega} + \frac{3}{4} \right)},
\end{eqnarray}
where
\begin{eqnarray}
a_{\mathrm{ho},\mu}^{(k)} = \sqrt{\frac{\hbar}{\mu^{(k)} \omega}}.
\end{eqnarray}
Since Fermi's regularized zero-range potential 
$V_{\mathrm{F,reg}}^{(k)}$ only acts at $r=0$,
the eigenstates of $H^{\mathrm{eff}}$ with non-vanishing
angular momentum $l_{\mathrm{eff}}$
do not feel the interaction and the corresponding eigenenergies
coincide with those of the 
non-interacting system.
Figure~\ref{fig_energyn2}
shows the relative eigenenergies $E_{\mathrm{rel}}(1,1)$ obtained by solving 
Eq.~(\ref{eq_busch}) for $k=aa$ as a function of $a_{\mathrm{ho}}/a^{(aa)}$
[in this case, $E_{\mathrm{eff}}=E_{\mathrm{rel}}(1,1)$
and $\mu^{(k)}=m/2$].
From Eq.~(\ref{eq_busch})
one finds at unitarity
$E_{\mathrm{unit,rel}}(1,1)=(2n_{\mathrm{eff}}+1/2) \hbar \omega$
for $s$-wave states.
The eigenenergies $E_{\mathrm{eff}}$
of the effective atom-dimer and dimer-dimer systems
can be obtained from Fig.~\ref{fig_energyn2} by scaling the
horizontal axis appropriately.

Within the effective two-particle model,
the 
relative energies $E_{\mathrm{rel}}(2,1)$ and $E_{\mathrm{rel}}(2,2)$
of the three- 
and four-fermion systems are
given by
$E_{\mathrm{rel}}(2,1)=E_{\mathrm{eff}} + E_{\mathrm{dimer}}$
and $E_{\mathrm{rel}}(2,2)=E_{\mathrm{eff}}+2 E_{\mathrm{dimer}}$,
where the second term on the right hand sides
accounts for the internal molecular binding energy
of the dimer(s) in the presence of the trap~\cite{stec07b,stec08}.
$E_{\mathrm{dimer}}$ is given by the lowest energy
solution of Eq.~(\ref{eq_busch}) with $k=aa$
and $\mu^{(aa)}=m/2$ 
(see solid line in Fig.~\ref{fig_energyn2}).
For $a^{(aa)} \ll a_{\mathrm{ho}}$, the size of the
dimer---given to first order by $a^{(aa)}$---is much smaller than the
trap size and
$E_{\mathrm{dimer}}$ approaches the free-space result
$E_{\mathrm{dimer}}^{\mathrm{free}} = -\hbar^2/[m (a^{(aa)})^2]$.
As $a^{(aa)}$ increases, the role of the confinement becomes
increasingly more important and the lowest energy solution 
of
Eq.~(\ref{eq_busch}) starts to deviate 
from $E_{\mathrm{dimer}}^{\mathrm{free}}$.
Expanding Eq.~(\ref{eq_busch}) about the non-interacting 
energies $(2n_{\mathrm{eff}}+3/2)\hbar \omega$, the $s$-wave energies $E_{\mathrm{eff}}$
can be approximated as~\cite{busc98}
\begin{eqnarray}
\label{eq_shiftdimer}
E_{\mathrm{eff}} \approx  \nonumber \\
\left(
2n_{\mathrm{eff}} +3/2 +
\frac{2 \Gamma(n_{\mathrm{eff}}+3/2)}{\sqrt{\pi} \Gamma(n_{\mathrm{eff}}+1)\Gamma(3/2)} 
\frac{a^{(k)}}{a_{\mathrm{ho},\mu}^{(k)}} \right)
\hbar \omega
\end{eqnarray}
for small $|a^{(k)}|$.
Alternatively, this result can 
be obtained by
treating the atom-dimer and dimer-dimer interactions within first order
perturbation theory.

Lastly, we discuss the angular momentum $L_{\mathrm{rel}}$
of the three- and four-fermion systems implied by the effective
model Hamiltonian $H^{\mathrm{eff}}$.
If $E_{\mathrm{eff}}$ is taken to be one of the positive energy solutions
of Eq.~(\ref{eq_busch}), 
%implying $l_{\mathrm{eff}}=0$,
then the total relative angular momentum
$L_{\mathrm{rel}}$ of the three- and four-fermion systems is $0$
and the states have natural parity, i.e., $\Pi_{\mathrm{rel}}=+1$.
If $E_{\mathrm{eff}}$ is taken to be 
an eigenenergy of $H^{\mathrm{eff}}$ with finite angular momentum $l_{\mathrm{eff}}$, i.e.,
$E_{\mathrm{eff}}=(2n_{\mathrm{eff}}+l_{\mathrm{eff}}+3/2)\hbar \omega$ with $n_{\mathrm{eff}}=0,1,\cdots$,
then the total relative 
angular momentum of the three- and four-fermion systems
is $L_{\mathrm{rel}}=l_{\mathrm{eff}}$ and the states have, as above, natural parity,
i.e., $\Pi_{\mathrm{rel}}=(-1)^{L_{\mathrm{rel}}}$.
These observations are used below to interpret 
Figs.~\ref{fig_energyn3scaled} and \ref{fig_energyn4scale}.

\subsection{Hyperspherical Coordinate Approach}
\label{sec_hyperspherical}
The hyperspherical framework serves two distinct
purposes in this paper. It allows for (i) the construction
of non-interacting wave functions with good quantum numbers
and (ii) the classification of the energy spectrum at unitarity. 
This section first treats the non-interacting Fermi gas 
using hyperspherical coordinates and then reviews how the
formalism, with some modifications, carries over to the infinitely
strongly-interacting unitary Fermi gas.

To construct the eigenfunctions of the
non-interacting Fermi gas,
we write
the many-body Hamiltonian $H^{\mathrm{ni}}$,
Eq.~(\ref{eq_hamni}), in
hyperspherical 
coordinates~\cite{aver89,lin95,bohn98,timo02,timo04,ripe05,wern06,ritt06,reviewgreen}.
We first separate off the center-of-mass vector
$\vec{R}_{\mathrm{cm}}$ and 
then
divide the remaining $3N-3$ coordinates into $3N-4$ hyperangles, collectively
denoted by 
$\vec{\Omega}$ (see below for their definition), 
and the hyperradius $R$, 
\begin{eqnarray}
\label{eq_rhyper}
R^2 = \frac{1}{N}\sum_{j=1}^{N} (\vec{r}_j-\vec{R}_{\mathrm{cm}})^2. 
\end{eqnarray}
Using these coordinates, the Hamiltonian 
$H^{\mathrm{ni}}$ can be written
as
\begin{eqnarray}
\label{eq_ni}
H^{\mathrm{ni}} = H^{\mathrm{cm}}  
-\frac{\hbar^2}{2M} \left( 
\frac{\partial^2}{\partial R^2} + 
\frac{3N-4}{R} \frac{\partial}{\partial R}\right) + \nonumber \\
\frac{\Lambda^2 }{2M R^2} 
+\frac{1}{2}M \omega^2 R^2,
\end{eqnarray}
where the center-of-mass Hamiltonian $H^{\mathrm{cm}}$
is given by
\begin{eqnarray}
H^{\mathrm{cm}}= \frac{-\hbar^2}{2M}\nabla_{\vec{R}_{\mathrm{cm}}}^2 +
\frac{1}{2}M \omega^2 R_{\mathrm{cm}}^2
\end{eqnarray}
and $M$ denotes the total mass of the system, i.e., $M=Nm$. 
In Eq.~(\ref{eq_ni}), 
$\Lambda$ denotes the so-called grand angular momentum 
operator~\cite{aver89}
that accounts for the kinetic energy associated with the hyperangles
$\vec{\Omega}$.
The eigenfunctions $\psi^{\mathrm{ni}}$ 
of the Hamiltonian $H^{\mathrm{ni}}$ separate
(see, e.g., Refs.~\cite{aver89,wern06,ritt06}),
\begin{eqnarray}
\label{eq_waveni}
\psi^{\mathrm{ni}}(\vec{r}_1,\cdots,\vec{r}_N) = \nonumber \\
G_{N_{\mathrm{cm}},L_{\mathrm{cm}},M_{\mathrm{cm}}}(\vec{R}_{\mathrm{cm}}) F_{q,\lambda}(R) 
\Phi_{\lambda,\chi}(\vec{\Omega}).
\end{eqnarray}
Here, the center-of-mass functions 
$G_{N_{\mathrm{cm}},L_{\mathrm{cm}},M_{\mathrm{cm}}}(\vec{R}_{\mathrm{cm}})$ are 
eigenfunctions of $H^{\mathrm{cm}}$, i.e., 
three-dimensional 
harmonic oscillator functions in the center-of-mass vector $\vec{R}_{\mathrm{cm}}$,
with eigenenergies $E_{\mathrm{cm}}=(2N_{\mathrm{cm}}+L_{\mathrm{cm}}+3/2)\hbar \omega$,
where $N_{\mathrm{cm}}=0,1,\cdots$, $L_{\mathrm{cm}}=0,1,\cdots$ and 
$M_{\mathrm{cm}}=-L_{\mathrm{cm}},-L_{\mathrm{cm}}+1,\cdots,L_{\mathrm{cm}}$.
The hyperspherical harmonics
$\Phi_{\lambda,\chi}(\vec{\Omega})$, or so-called channel functions,
are eigenfunctions of the
operator $\Lambda^2$~\cite{aver89},
\begin{eqnarray}
\label{eq_wavechannel}
\Lambda^2 \Phi_{\lambda,\chi}(\vec{\Omega}) = 
\hbar^2 \lambda(\lambda + 3N-5)
\Phi_{\lambda,\chi}(\vec{\Omega}),
\end{eqnarray}
where $\lambda$ can take
the values $0,1,2,\cdots$. 
The quantum number $\chi$ denotes the degeneracy for each 
$\lambda$~\cite{aver89},
\begin{eqnarray}
\label{eq_chi}
\chi=\frac{(3N+2 \lambda-5)(3N+\lambda-6)!}{\lambda! (3N-5)}.
\end{eqnarray}
In deriving 
Eq.~(\ref{eq_chi}),
no symmetry constraints have been enforced.
Below, we discuss the 
construction of the hyperspherical harmonics
and the reduction of the degeneracy $\chi$
due to symmetry constraints for the three- and four-fermion systems.
Since the center-of-mass coordinates and the hyperradius are unchanged
under the exchange of $\vec{r}_j$ and $\vec{r}_k$ ($j,k=1,\cdots,N$),
the symmetry-constraints only effect 
the $\Phi_{\lambda,\chi}(\vec{\Omega})$ 
and neither $G_{N_{\mathrm{cm}},L_{\mathrm{cm}},M_{\mathrm{cm}}}(\vec{R}_{\mathrm{cm}})$ 
nor $F_{q,\lambda}(R)$.

Plugging Eq.~(\ref{eq_waveni}) into 
the Schr\"odinger equation 
$H^{\mathrm{ni}}\psi^{\mathrm{ni}}=E_{\mathrm{ni}}(N_{\uparrow},N_{\downarrow})\psi^{\mathrm{ni}}$
and
dividing out the center-of-mass and hyperangular
contributions,
we obtain an
effective hyperradial Schr\"odinger equation~\cite{wern06,ritt06},
\begin{eqnarray}
\label{eq_hyperradial}
\left( 
-\frac{\hbar^2}{2M} \frac{\partial^2}{\partial R^2} +
\frac{\hbar^2K_{\mathrm{ni}}(K_{\mathrm{ni}}+1)}{2MR^2} 
+\frac{1}{2}M \omega^2 R^2 \right) \bar{\mathrm{F}}_{q,\lambda}(R)  \nonumber \\
=\big[E_{\mathrm{ni}}(N_{\uparrow},N_{\downarrow})-E_{\mathrm{cm}}\big] \bar{\mathrm{F}}_{q,\lambda}(R),
\end{eqnarray}
where 
\begin{eqnarray}
\bar{\mathrm{F}}_{q,\lambda}(R) = R^{(3N-4)/2}F_{q,\lambda}(R)
\end{eqnarray}
and
\begin{eqnarray}
\label{eq_capk}
K_{\mathrm{ni}} = \lambda + \frac{3N-6}{2}.
\end{eqnarray}
Noticing that 
the effective hyperradial Schr\"odinger
equation, Eq.~(\ref{eq_hyperradial}), is 
formally identical to the Schr\"odinger equation
for the three-dimensional harmonic oscillator with 
angular momentum $K_{\mathrm{ni}}$~\cite{wern06,ritt06},
the eigenenergies $E_{\mathrm{ni,rel}}$, 
$E_{\mathrm{ni,rel}}(N_{\uparrow},N_{\downarrow})=E_{\mathrm{ni}}(N_{\uparrow},N_{\downarrow})-E_{\mathrm{cm}}$, and 
the corresponding eigenfunctions $\bar{\mathrm{F}}_{q,\lambda}(R)$ are
readily written down,
\begin{eqnarray}
\label{eq_energyrel}
E_{\mathrm{ni,rel}} = \left( 2q + K_{\mathrm{ni}} + \frac{3}{2} \right)\hbar \omega
\end{eqnarray}
with $q=0,1,\cdots$ 
and
\begin{eqnarray}
\label{eq_wavehyperradial}
\bar{\mathrm{F}}_{q,\lambda}(R)=\nonumber \\ 
N_{qK_{\mathrm{ni}}} R^{K_{\mathrm{ni}}+1}
\exp\left( -\frac{R^2}{2a_M^2}\right) 
L_q^{(K_{\mathrm{ni}}+1/2)}\left( \frac{R^2}{a_M^2} \right).
\end{eqnarray}
The normalization constant $N_{qK_{\mathrm{ni}}}$ is
chosen such that
\begin{eqnarray}
\label{eq_normhyperradial1}
\int_0^{\infty} |\bar{\mathrm{F}}_{q,\lambda}(R)|^2 dR = 1,
\end{eqnarray}
leading to
\begin{eqnarray}
\label{eq_normhyperradial2}
N_{q K_{\mathrm{ni}}}=\sqrt{\frac{2^{K_{\mathrm{ni}}+2}}{(2K_{\mathrm{ni}}+1)!! \sqrt{\pi}
L_q^{(K_{\mathrm{ni}}+1/2)}(0)a_{\mathrm{ho},M}^{2K_{\mathrm{ni}}+3}}},
\end{eqnarray}
where $L_q^{(K_{\mathrm{ni}}+1/2)}$ 
denotes
the associated Laguerre polynomial.
The harmonic oscillator 
length $a_{\mathrm{ho},M}$, $a_{\mathrm{ho},M}=\sqrt{\hbar/(M\omega)}$, 
can be interpreted as being associated with an effective mass $M$ particle
that moves along the hyperradial coordinate $R$.
The quantity $K_{\mathrm{ni}}$ depends on $\lambda$ and
can be thought of as an effective angular momentum quantum number;
it should be noted, though, that the
$K_{\mathrm{ni}}$ are, in general, neither equal to the
total angular momentum $L_{tot}$ nor equal to
the relative angular momentum $L_{\mathrm{rel}}$ of the two-component Fermi gas.

The explicit 
construction of the hyperspherical harmonics 
$\Phi_{\lambda,\chi}(\vec{\Omega})$
requires the hyperangles $\vec{\Omega}$ to be specified.
The hyperangles $\vec{\Omega}$ can be defined in many different
ways and here we employ definitions that allow for a 
straightforward anti-symmetrization of the 
$\Phi_{\lambda,\chi}(\vec{\Omega})$.
To this end, we introduce a set
of mass scaled Jacobi vectors $\vec{u}_i$, $i=1,\cdots,N-1$, 
where~\cite{aver89,lin95}
\begin{eqnarray}
\vec{u}_i = \sqrt{\frac{M}{\mu_i}} \vec{\rho}_i.
\end{eqnarray}
Table~\ref{tab_jacobifermi}
\begin{table}
\caption{Definition of
the Jacobi vectors $\vec{\rho}_i$ and the associated 
reduced masses $\mu_i$
used in our construction
of the hyperspherical harmonics $\Phi_{\lambda,\chi}$
for the two-component
equal-mass atomic Fermi gas with $(N_{\uparrow},N_{\downarrow})=(2,1)$
and $(2,2)$. 
}
\begin{ruledtabular}
\begin{tabular}{l|ccc|ccc}
  $(N_{\uparrow},N_{\downarrow})$ & 
$\vec{\rho}_1$ & $\vec{\rho}_2$ &  
$\vec{\rho}_3$ & 
$\mu_1$ & $\mu_2$ & $\mu_3$ 
\\
\hline
$(2,1)$ & $\vec{r}_1-\vec{r}_2$ & $\frac{\vec{r}_1+\vec{r}_2}{2}-\vec{r}_3$ 
& & $\frac{m}{2}$ & $\frac{2m}{3}$ &  \\
$(2,2)$ & $\vec{r}_1-\vec{r}_2$ & $\vec{r}_3-\vec{r}_4$ & 
$\frac{\vec{r}_1+\vec{r}_2}{2}-\frac{\vec{r}_3+\vec{r}_4}{2}$
&  $\frac{m}{2}$ & $\frac{m}{2}$ & $m$   
\end{tabular}
\end{ruledtabular}
\label{tab_jacobifermi}
\end{table}
lists the Jacobi coordinates $\vec{\rho}_i$ and
the associated reduced masses $\mu_i$
employed 
in this paper
for the treatment of atomic two-component equal-mass
three- and four-fermion systems.
These Jacobi vectors 
have particularly
convenient properties under the exchange of identical fermions.
For the $(N_{\uparrow},N_{\downarrow})=(2,1)$ system,
the Jacobi vector $\vec{\rho}_1$ changes sign under the exchange of the 
two spin-up fermions while $\vec{\rho}_2$ remains unchanged. 
For the
$(N_{\uparrow},N_{\downarrow})=(2,2)$
system, 
the exchange of the two spin-up fermions leads to
a sign change of $\vec{\rho}_1$ while $\vec{\rho}_2$ and $\vec{\rho}_3$ 
remain unchanged,
the exchange of the two spin-down fermions leads to
a sign change of $\vec{\rho}_2$ while $\vec{\rho}_1$ and $\vec{\rho}_3$ 
remain unchanged,
and the simultaneous exchange of the 
two
spin-up fermions and the two spin-down fermions leads to
a sign change of $\vec{\rho}_1$ and $\vec{\rho}_2$ while
$\vec{\rho}_3$ remains unchanged.
These properties of the Jacobi vectors make the construction
of properly anti-symmetrized hyperspherical
harmonics $\Phi_{\lambda,\chi}(\vec{\Omega})$ for
the three- and four-fermion systems comparatively simple.
In terms of the mass-scaled Jacobi vectors $\vec{u}_i$,
the $3N-4$ hyperspherical angles $\vec{\Omega}$ are
defined
as $\vec{\Omega}=(\vec{u}_1/R,\cdots,\vec{u}_{N-1}/R)$
and
the hyperradius $R$, Eq.~(\ref{eq_rhyper}), can
be rewritten as
$R^2=\sum_{i=1}^{N-1} \vec{u}_i^2$~\cite{aver89,lin95}.

Following Avery~\cite{aver89},
we construct a complete set
of hyperspherical harmonics 
$\Phi_{\lambda,\chi}(\vec{\Omega})$,
which are simultaneous eigenfunctions
of the operators $\Lambda^2$, $L^2_{\mathrm{rel}}$, 
$L_{\mathrm{rel},z}$ and $\Pi_{\mathrm{rel}}$.
Although the explicit 
functional forms of the $\Phi_{\lambda,\chi}(\vec{\Omega})$
are needed in our perturbative treatment, 
we restrict ourselves here to summarizing the degeneracies
of the non-interacting eigenfunctions for the
$(N_{\uparrow},N_{\downarrow})=(2,1)$ and $(2,2)$
systems (see Tables~\ref{tab_hhn3} 
\begin{table}
\caption{Characterization of
the hyperspherical harmonics 
$\Phi_{\lambda,\chi}(\vec{\Omega})$
for the $(N_{\uparrow},N_{\downarrow})=(2,1)$ system with $\lambda \le 5$.
In determining $\chi$,
only hyperspherical harmonics that change sign under the exchange
of the two spin-up atoms are counted.
}
\begin{ruledtabular}
\begin{tabular}{ccccc}
$\lambda$ & $\chi$ & $K_{\mathrm{ni}}$ & $L_{\mathrm{rel}}$ & ${\Pi}_{\mathrm{rel}}$ \\
\hline
$1$ & 3 & 5/2 & 1 & $-1$ \\ \hline
2 & 5 & 7/2 & 2 & $+1$ \\
2 & 3 & 7/2 & 1 & $+1$ \\
2 & 1 & 7/2 & 0 & $+1$ \\ \hline
3 & 14 & 9/2 & 3 & $-1$ \\
3 & 5 & 9/2 & 2 & $-1$ \\
3 & 6 & 9/2 & 1 & $-1$ \\ \hline
4 & 18 & 11/2 & 4 & $+1$ \\
4 & 14 & 11/2 & 3 & $+1$ \\
4 & 15 & 11/2 & 2 & $+1$ \\
4 & 3 & 11/2 & 1 & $+1$ \\
4 & 1 & 11/2 & 0 & $+1$ \\ \hline
5 & 33 & 13/2 & 5 & $-1$ \\
5 & 18 & 13/2 & 4 & $-1$ \\
5 & 28 & 13/2 & 3 & $-1$ \\
5 & 10 & 13/2 & 2 & $-1$ \\
5 & 9 & 13/2 & 1 & $-1$ \\ 
\end{tabular}
\end{ruledtabular}
\label{tab_hhn3}
\end{table} 
and 
\begin{table}
\caption{Characterization of
the hyperspherical harmonics 
$\Phi_{\lambda,\chi}(\vec{\Omega})$
for the $(N_{\uparrow},N_{\downarrow})=(2,2)$ system with $\lambda \le 5$.
In determining $\chi$,
only hyperspherical harmonics that change sign under the exchange
of the two spin-up atoms 
and the two spin-down atoms 
are counted.
}
\begin{ruledtabular}
\begin{tabular}{ccccc}
$\lambda$ & $\chi$ & $K_{\mathrm{ni}}$ & $L_{\mathrm{rel}}$ & ${\Pi}_{\mathrm{rel}}$ \\
\hline
$2$ & 5 & 5 & 2 & $+1$ \\
$2$ & 3 & 5 & 1 & $+1$ \\
$2$ & 1 & 5 & 0 & $+1$ \\ \hline
$3$ & 7 & 6 & 3 & $-1$ \\
$3$ & 10 & 6 & 2 & $-1$ \\
$3$ & 9 & 6 & 1 & $-1$ \\
$3$ & 1 & 6 & 0 & $-1$ \\ \hline
$4$ & 27 & 7 & 4 & $+1$ \\
$4$ & 28 & 7 & 3 & $+1$ \\
$4$ & 35 & 7 & 2 & $+1$ \\
$4$ & 12 & 7 & 1 & $+1$ \\
$4$ & 3 & 7 & 0 & $+1$ \\ \hline
$5$ & 33 & 8 & 5 & $-1$ \\
$5$ & 54 & 8 & 4 & $-1$ \\
$5$ & 77 & 8 & 3 & $-1$ \\
$5$ & 50 & 8 & 2 & $-1$ \\
$5$ & 27 & 8 & 1 & $-1$ \\
$5$ & 2 & 8 & 0 & $-1$ \\
\end{tabular}
\end{ruledtabular}
\label{tab_hhn4}
\end{table} 
\ref{tab_hhn4}).

Table~\ref{tab_hhn3} shows the degeneracies $\chi$
and quantum numbers of the hyperspherical harmonics
for the $(N_{\uparrow},N_{\downarrow})=(2,1)$
system with $\lambda$ up to $5$; in constructing Table~\ref{tab_hhn3},
only hyperspherical harmonics that
change sign 
under the exchange of
the two spin-up atoms have been counted.
This symmetry constraint reduces the degeneracy of each
$\lambda$ manifold tremendously.
Equation~(\ref{eq_chi})---applicable to a system without
symmetry constraints---gives 1, 6, 20, and 50 for
$\lambda=0,1,2$ and 3, while
Table~\ref{tab_hhn3} shows that the 
degeneracies are reduced to 0, 3, 9 and 25.
Table~\ref{tab_hhn3} 
can be readily constructed by considering the
angular momentum
operators $\vec{l}_{1}$ and $\vec{l}_{2}$
associated with  the Jacobi vectors $\vec{\rho}_1$
and $\vec{\rho}_2$, and by
taking into account that $\vec{l}_1$ and $\vec{l}_2$
couple to $\vec{L}_{\mathrm{rel}}$~\cite{aver89}.
Since $\vec{\rho}_1$ 
is the Jacobi vector that connects the two spin-up fermions,
$l_1$ can only take odd values; $l_2$, in contrast, is not
restricted by symmetry constraints, implying $l_2=0,1,\cdots$.
For a given $\lambda$, 
%$l_1+l_2$ can take the values $\lambda,\lambda-2,\cdots,0$
%for even $\lambda$ and
%$\lambda,\lambda-2,\cdots,1$ for odd $\lambda$~\cite{aver89,review}.
the allowed $(l_1,l_2)$ combinations are determined
by
$\lambda=l_1+l_2+2p$, where $p=0,1,\cdots$~\cite{aver89,reviewgreen}.
Since the
$(l_1,l_2)=(0,0)$ combination is symmetry-forbidden,
the smallest allowed $\lambda$
value is $1$.
For $\lambda=1$, the only possible $(l_1,l_2)$ combination is 
$(1,0)$, resulting in $L_{\mathrm{rel}}=1$, $\Pi_{\mathrm{rel}}=(-1)^{l_1+l_2}=-1$
and a degeneracy of $\chi=3$ (corresponding to three
different projection quantum numbers $M_L$).
For $\lambda=2$, the only possible $(l_1,l_2)$
combination is $(1,1)$, leading to  
$L_{\mathrm{rel}}=0,1,2$ and $\Pi_{\mathrm{rel}}=+1$.
The degeneracy $\chi$ is 9 (1, 3 and 5 states for $L_{\mathrm{rel}}=0$, 1 
and 2, respectively).
For $\lambda=3$,
the allowed $(l_1,l_2)$ combinations are $(3,0)$,
$(1,2)$ and $(1,0)$, leading to $L_{\mathrm{rel}}=3$ (7 states),
$L_{\mathrm{rel}}=3,2,1$ (15 states) and $L_{\mathrm{rel}}=1$
(3 states), respectively; thus, the degeneracy
$\chi$ is 25.
Following this reasoning, the remaining entries in Table~\ref{tab_hhn3}
can be verified.

Table~\ref{tab_hhn4} summarizes the degeneracies and quantum numbers for the
$(N_{\uparrow},N_{\downarrow})=(2,2)$ system.
Similarly to the three-fermion case, Table~\ref{tab_hhn4} 
is constructed by
realizing that the angular momentum quantum numbers $l_1$ and $l_2$ 
associated with the Jacobi vectors $\vec{\rho}_1$ and 
$\vec{\rho}_2$ can only take odd values
and that $l_3$, where 
$l_3$ denotes the angular momentum quantum number
associated with the Jacobi vector $\vec{\rho}_3$,
can take any value.
For a given $\lambda$,
the allowed $(l_1,l_2,l_3)$ combinations are determined
by
$\lambda=l_1+l_2+l_3+2p+2q$, where 
$p,q=0,1,\cdots$~\cite{aver89,reviewgreen}.
%$l_1+l_2$ can take the values
%$\tilde{\lambda}$, where 
%$\tilde{\lambda}=\lambda,\lambda-2,\cdots,0$
%for $\lambda$ even
%and
%$\tilde{\lambda}=\lambda,\lambda-2,\cdots,1$
%for $\lambda$ odd.
%For each $\lambda$ and $\tilde{\lambda}$, 
%in turn, $\tilde{\lambda}+l_3$ can take the values
%$\lambda,\lambda-2,\cdots,0$ for $\lambda$ even
%and
%$\lambda,\lambda-2,\cdots,1$ for $\lambda$ odd.
Since both $l_1$ and $l_2$ have to be odd, the smallest allowed
$\lambda$ value is 2. In this case, 
%$\tilde{\lambda}=2$ 
$l_1=l_2=1$
and $l_3=0$.
The $\lambda=2$ manifold thus consists of nine states
[$\vec{l}_1$, $\vec{l}_2$ and $\vec{l}_3$ can couple so that $L_{\mathrm{rel}}=2$
(5 states), $1$ (3 states) and 0 (1 state)].
For $\lambda=3$, the only possibility is
$(l_1,l_2,l_3)=(1,1,1)$, implying $27$
states.
The angular momenta corresponding
to these 27 states can be obtained by first coupling $\vec{l}_1$
and $\vec{l}_2$ to an intermediate
angular momentum vector with
quantum number $2$, 1 or 0, and then coupling the intermediate
angular momentum vector and $\vec{l}_3$ to obtain $\vec{L}_{\mathrm{rel}}$.
The higher $\lambda$ manifolds are treated
following the same scheme.

Knowing the allowed $\lambda$ and $\chi$ values, the degeneracy 
$g_{\mathrm{ni,rel}}$ of a given 
relative energy $E_{\mathrm{ni,rel}}(N_{\uparrow},N_{\downarrow})$ of the
non-interacting trapped Fermi gas can be easily determined using 
Eq.~(\ref{eq_capk}) and Eq.~(\ref{eq_energyrel}). 
These degeneracies are summarized in the second
column of Tables~\ref{tab_shift3}
\begin{table}
\caption{Coefficients $c^{(1)}$ 
for Fermi gas with $(N_{\uparrow},N_{\downarrow})=(2,1)$.
The $c^{(1)}$ are defined through 
$E^{(1)}= c^{(1)} (2 \pi)^{-1/2} \hbar \omega a^{(aa)}/a_{\mathrm{ho}}$.}
\begin{ruledtabular}
\begin{tabular}{ccccc}
$E_{\mathrm{ni,rel}}/(\hbar \omega)$ & $g_{\mathrm{ni,rel}}$ & $L_{\mathrm{rel}}$ & ${\Pi}_{\mathrm{rel}}$ & $c^{(1)}$ \\
\hline
$4$ & 3 & 1 & $-1$ & 3 \\ \hline
$5$ & 5 & 2 & $+1$ & 3/2 \\
$5$ & 3 & 1 & $+1$ & 0 \\
$5$ & 1 & 0 & $+1$ & 15/4 \\ \hline
$6$ & 7 & 3 & $-1$ & 9/4 \\
$6$ & 7 & 3 & $-1$ & 0 \\
$6$ & 5 & 2 & $-1$ & 0 \\
$6$ & 3 & 1 & $-1$ & $\frac{3}{16}(13+\sqrt{41})$ \\
$6$ & 3 & 1 & $-1$ & $\frac{3}{16}(13-\sqrt{41})$ \\
$6$ & 3 & 1 & $-1$ & 0 
\end{tabular}
\end{ruledtabular}
\label{tab_shift3}
\end{table} 
and \ref{tab_shift4} for the three- and four-fermion
\begin{table}
\caption{Coefficients $c^{(1)}$
for Fermi gas with $(N_{\uparrow},N_{\downarrow})=(2,2)$.
The $c^{(1)}$ are defined through 
$E^{(1)}= c^{(1)} (2 \pi)^{-1/2} \hbar \omega a^{(aa)}/a_{\mathrm{ho}}$.}
\begin{ruledtabular}
\begin{tabular}{ccccc}
$E_{\mathrm{ni,rel}}/(\hbar \omega)$ & $g_{\mathrm{ni,rel}}$ & $L_{\mathrm{rel}}$ & ${\Pi}_{\mathrm{rel}}$ & $c^{(1)}$ \\
\hline
13/2	& 5 & 2  & $+1$ & 5	 \\ 
13/2	& 3 & 1  & $+1$ & 4	 \\ 
13/2	& 1 & 0  & $+1$ & 13/2	 \\ 		\hline
15/2	& 7 & 3	 & $-1$ & 7/2			 \\ 
15/2	& 5 & 2  & $-1$ & 3			 \\ 	
15/2	& 5 & 2	 & $-1$ & 2			 \\ 
15/2	& 3 & 1  & $-1$ & 5 	 \\ 			
15/2	& 3 & 1	 & $-1$ & $\frac{1}{8}(29+\sqrt{41})$ \\ 
15/2	& 3 & 1	 & $-1$ & $\frac{1}{8}(29-\sqrt{41})$ \\ 
15/2	& 1 & 0  & $-1$ & 0	 \\ \hline
17/2	& 9 & 4	 & $+1$ & $\frac{1}{8}(27+\sqrt{73})$	 \\ 
17/2	& 9 & 4	 & $+1$ & 5/2				 \\ 
17/2	& 9 & 4	 & $+1$ & $\frac{1}{8}(27-\sqrt{73})$	 \\ 
17/2	& 7 & 3	 & $+1$ & $\frac{1}{4}(9+\sqrt{17})$ \\ 	
17/2	& 7 & 3	 & $+1$ & 3			 \\ 	
17/2	& 7 & 3	 & $+1$ & $\frac{1}{4}(9-\sqrt{17})$	 \\ 
17/2	& 7 & 3	 & $+1$ & 0				 \\ 
17/2	& 5 & 2	 & $+1$ & 5.89252 \\ 			
17/2	& 5 & 2	 & $+1$ & 5.31030	 \\ 		
17/2	& 5 & 2	 & $+1$ & 4.61321 \\ 			
17/2	& 5 & 2	 & $+1$ & 3.21783 \\ 			
17/2	& 5 & 2	 & $+1$ & 3.21549 \\ 			
17/2	& 5 & 2	 & $+1$ & 1.92129	 \\ 		
17/2	& 5 & 2	 & $+1$ & 1.45435 \\ 			
17/2	& 5 & 2	 & $+1$ & 0		 \\ 
17/2	& 3 & 1	 & $+1$ & 4.50566 \\ 			
17/2	& 3 & 1	 & $+1$ & 3	 \\ 			
17/2	& 3 & 1	 & $+1$ & 1.73167 \\ 			
17/2	& 3 & 1	 & $+1$ & 0.512668 \\ 			
17/2	& 3 & 1	 & $+1$ & 0	 \\ 			
17/2	& 1 & 0	 & $+1$ & 7.40848		 \\ 
17/2	& 1 & 0	 & $+1$ & 6.98138	 \\ 
17/2	& 1 & 0	 & $+1$ & 15/4		 	 \\ 	
17/2	& 1 & 0	 & $+1$ & 2.89139 
\end{tabular}
\end{ruledtabular}
\label{tab_shift4}
\end{table} 
systems,
respectively.
Alternatively~\cite{timo02,ritt06},
the relative energy $E_{\mathrm{ni,rel}}(N_{\uparrow},N_{\downarrow})$
of the non-interacting three- and four-fermion systems
can be written as
$E_{\mathrm{ni,rel}}(N_{\uparrow},N_{\downarrow})=\sum_{j=1}^{N-1}(2n_j+l_j+3/2) \hbar \omega$,
where $n_j=0,1,\cdots$ and where
the allowed angular momentum quantum numbers $l_j$ are
%,
%as discussed above, 
%constrained 
determined 
by the symmetry
requirements (see above).
% and where 
Counting the possible combinations of $l_j$ and $n_j$ values
and taking the $(2l_j+1)$ degeneracy associated with each $l_j$
into account,
gives the same results as those reported in the second column of
Tables~\ref{tab_shift3} and \ref{tab_shift4},
and also allows---using Eqs.~(\ref{eq_capk})
and (\ref{eq_energyrel})---for an independent
determination of the $\lambda$
and $\chi$ values given 
in the first two columns of Tables~\ref{tab_hhn3} and \ref{tab_hhn4}.

So far, we have discussed the hyperspherical framework
for the non-interacting two-component Fermi gas.
We now review the modifications needed when applying this 
framework
to the infinitely strongly-interacting unitary gas
with zero-range two-body interactions.
The zero-range 
two-body potential with
infinite $a^{(aa)}$ does not establish
a meaningful length scale, leaving the oscillator length 
$a_{\mathrm{ho}}$ as the only length scale in the problem.
Using scale invariance arguments,
it has been shown~\cite{wern06} that a diverging $s$-wave
scattering length $a^{(aa)}$ implies that the wave function
$\psi^{\mathrm{unit}}(\vec{r}_1,\cdots,\vec{r}_N)$ at unitarity
separates in the same way as that of the non-interacting 
system [see Eq.~(\ref{eq_waveni})].
It follows that Eq.~(\ref{eq_hyperradial}) applies
not only to the non-interacting gas but also to the
unitary gas if $K_{\mathrm{ni}}$
is replaced by $K_{\mathrm{unit}}$
and if $\lambda$ is reinterpreted as 
the eigenvalue of the hyperangular eigenequation
that takes the two-body interactions into account.
In the following, we use $K_{\mathrm{unit}}$ to 
denote the effective angular momentum of the 
unitary gas.
Note that $K_{\mathrm{unit}}$
depends on the eigenvalue of the hyperangular eigenequation, i.e.,
there exists a $K_{\mathrm{unit}}$ for each channel function
$\Phi_{\lambda,\chi}(\vec{\Omega})$; for notational convenience, 
we do not explicitly indicate the
dependence of $K_{\mathrm{unit}}$ on the hyperangular 
quantum numbers.

The coefficients $K_{\mathrm{unit}}$
have beeen obtained
for all states of the three-fermion system~\cite{wern06a}
(see also Refs.~\cite{efim71,efim73,dinc05} 
for earlier work)
and for the lowest 
20 states with $(L_{\mathrm{rel}},\Pi_{\mathrm{rel}})=(0,+1)$ 
of the four-fermion system~\cite{stec09}
by solving the hyperangular Schr\"odinger equation
that includes the two-body interactions (see also Ref.~\cite{blum07}).
The relative eigenenergies $E_{\mathrm{unit,rel}}$
of the
unitary gas are,
similarly to the non-interacting case, given by~\cite{wern06} 
\begin{eqnarray}
\label{eq_energyunit}
E_{\mathrm{unit,rel}}=(2q+K_{\mathrm{unit}}+3/2) \hbar \omega,
\end{eqnarray}
and
Eqs.~(\ref{eq_wavehyperradial})-(\ref{eq_normhyperradial2})
remain valid if $K_{\mathrm{ni}}$ is replaced by $K_{\mathrm{unit}}$
(and if $\lambda$ is reinterpreted as discussed above).
%While the values of $K_{\mathrm{ni}}$ for the non-interacting 
%gas are known,
%the
%values of $K_{\mathrm{unit}}$ for the unitary gas are only known
%for a subset of states of the four-fermion system~\cite{blum07,stec09}
%(see above).
In Sec.~\ref{sec_results}, we determine a number of $K_{\mathrm{unit}}$ coefficients
by solving the full relative Schr\"odinger equation
of the four-fermion system 
for various 
$L_{\mathrm{rel}}>0$ 
and by then comparing the resulting energy with 
the right hand side of Eq.~(\ref{eq_energyunit}).
Lastly,
we note that Eq.~(\ref{eq_energyunit}) implies 
that the excitation spectrum of
the
trapped unitary two-component Fermi gas
contains ladders of excitation frequencies that are integer multiples 
of $2 \hbar \omega$,
independent of the actual values of $K_{\mathrm{unit}}$~\cite{tan04,cast04,wern06}.

\subsection{Numerical treatment of the three-fermion system:
Lippmann-Schwinger equation}
\label{sec_threebody}
The trapped three-fermion problem with zero-range interactions 
and arbitrary $s$-wave scattering length $a^{(aa)}$ has been
solved using a number of different semi-analytical and numerical 
approaches~\cite{kest07,stet07,stec08}.
Here, we replace the regularized
zero-range pseudo-potential $V_{\mathrm{F,reg}}^{(aa)}(\vec{r})$,
which describes the interactions between atoms with opposite spins, 
by the corresponding Bethe-Peierls boundary condition
and employ an approach 
developed by
Kestner and Duan~\cite{kest07} that is based on the Lippmann-Schwinger
equation.
This approach reduces the three-body problem 
to solving a set of coupled equations~\cite{kest07},
\begin{eqnarray}
\label{eq_kestner}
\frac{2 
\Gamma \left(-\nu_j\right)}
{\Gamma\left( -\nu_j-\frac{1}{2}\right)} 
c_j + \sum_{k=1}^{B} d_{jk} c_k = 
\left(
\frac{a_{\mathrm{ho},\mu}^{(aa)}}{a^{(aa)}}
\right) c_j,
\end{eqnarray}
for the eigenvector $\vec{c}$,
$\vec{c}=(c_1,\cdots,c_B)$, and the eigenvalue
$(a^{(aa)}/a_{\mathrm{ho},\mu}^{(aa)})^{-1}$.
In Eq.~(\ref{eq_kestner}), the $\nu_j$ denote non-integer quantum 
numbers that depend on $E_{\mathrm{rel}}(2,1)$
and the $d_{jk}$ dimensionless
matrix elements. Their definitions are given in
Refs.~\cite{kest07,liu09}.
%In Eq.~(\ref{eq_kestner}), the non-integer quantum 
%numbers $\nu_j$ are defined as
%\begin{eqnarray}
%\nu_j=\frac{E_{\mathrm{rel}}(2,1)-(2j+L_{\mathrm{rel}}+3)\hbar \omega}{2 \hbar \omega}.
%\end{eqnarray}
%The dimensionless matrix elements
%$d_{jk}$
%are given by
%\begin{eqnarray}
%\label{eq_kestner2}
%d_{jk} = \frac{(-1)^{L_{\mathrm{rel}}}}{\sqrt{\pi}}
%\int_0^{\infty} R_{jL_{\mathrm{rel}}}(x) R_{kL_{\mathrm{rel}}}(x/2) 
%\times \nonumber \\
%\psi_{{\mathrm{tb}},\nu_k}(\sqrt{3}x/2)
%x^2 dx,
%\end{eqnarray}
%where 
%$R_{j L_{\mathrm{rel}}}(x)$ denotes
%the radial part of the three-dimensional harmonic oscillator functions,
%\begin{eqnarray}
%R_{jL_{\mathrm{rel}}}(x)=
%N_{jL_{\mathrm{rel}}} x^{L_{\mathrm{rel}}} \exp\left( -\frac{x^2}{2} \right) \times 
%\nonumber \\
%M\left(-j,L_{\mathrm{rel}}+\frac{3}{2},x^2 \right)
%\end{eqnarray}
%with
%\begin{eqnarray}
%N_{jL_{\mathrm{rel}}}=\sqrt{\frac{2^{L_{\mathrm{rel}}+2} \Gamma(L_{\mathrm{rel}}+n+3/2)}
%{\sqrt{\pi} j! (2L_{\mathrm{rel}}+1)!! \Gamma(L_{\mathrm{rel}}+3/2)}}
%\end{eqnarray}
%and where
%$\psi_{{\mathrm{tb}},\nu_k}(x)$ denotes the unnormalized $s$-wave two-body wave function
%for zero-range interactions,
%\begin{eqnarray}
%\psi_{{\mathrm{tb}},\nu_k}(x)= \Gamma(-\nu_k) \exp\left( \frac{-x^2}{2}\right) 
%U\left(-\nu_k,\frac{3}{2},x^2\right);
%\end{eqnarray}
%here, $M$ and $U$ denote 
%confluent hypergeometric functions.

In the $B \rightarrow \infty$ limit,
Eq.~(\ref{eq_kestner}) gives the exact three-fermion energy spectrum.
For each $E_{\mathrm{rel}}(2,1)$, there exist multiple $\vec{c}$ 
and $a^{(aa)}$
that solve Eq.~(\ref{eq_kestner}).
Thus, Eq.~(\ref{eq_kestner}) can be interpreted as a matrix equation with
eigenvector matrix $\underline{c}=(\vec{c}_1,\cdots,\vec{c}_B)$
and eigenvalue vector $((a^{(aa)})_1,\cdots,(a^{(aa)})_B)$.
The solutions obtained by solving Eq.~(\ref{eq_kestner})
belong to three-fermion states with natural parity. 
For the three-fermion system,
unnatural parity
states are not
affected by the $s$-wave zero-range interactions and coincide with those
of the non-interacting system.
%
%If we 
%consider Eq.~(\ref{eq_kestner}) with $d_{jk}=0$, then
%only 
%one of the two interactions 
%between spin-up and spin-down atoms is accounted for and the 
%three-body solution reduces
%to that of an $s$-wave interacting pair and a free atom. The fact that
%the second spin-up atom also interacts with the spin-down atom
%is reflected
%by the non-vanishing
%``exchange matrix elements'' $d_{jk}$ in Eq.~(\ref{eq_kestner}).
%The 
%approach outlined here
%Solving Eq.~(\ref{eq_kestner})
%allows for the determination of the three-body spectrum
%with moderate numerical efforts. 
For each $L_{\mathrm{rel}}$,
we solve the matrix problem for different $B$,
$B \le 50$. 
For positive $E_{\mathrm{rel}}(2,1)$, our results presented 
in Sec.~\ref{sec_resultsn3} are obtained using $B=50$.
For negative $E_{\mathrm{rel}}(2,1)$, we use somewhat smaller $B$ values;
we have checked through extrapolation to the
$B \rightarrow \infty$ limit that the three-fermion energies
obtained in this manner are highly accurate.
We find, e.g., that the eigenenergies at unitarity obtained
by the numerical 
approach based on the Lippmann-Schwinger equation~\cite{kest07} agree to
better than 0.01\% with those
obtained 
by solving the
transcendental equation derived by Werner and Castin~\cite{wern06a}.

\subsection{Numerical treatment of the four-fermion system:
Stochastic variational approach}
\label{sec_stochastic}
To determine the energy spectrum of
two-component Fermi gases with $(N_{\uparrow},N_{\downarrow})=(2,2)$
under spherically
symmetric harmonic confinement,
we employ the stochastic variational 
approach~\cite{varg95,varg01,cgbook,sore05,stec07c,stec07b,stec08}.
Our implementation
separates off the center-of-mass degrees of freedom $\vec{R}_{\mathrm{cm}}$,
defines a set of $N-1$ Jacobi coordinates 
$\vec{x}$, $\vec{x}=(\vec{\rho}_1,\cdots,\vec{\rho}_{N-1})$,
and expands the relative wave function $\psi_{\mathrm{rel}}(\vec{x})$
in terms of the basis functions
$\varphi_k(\vec{x})$,
\begin{eqnarray}
\label{eq_basis}
\psi_{\mathrm{rel}}(\vec{x}) = \sum_{k=1}^B 
{\cal{A}} 
\left[ 
c_k 
\varphi_k(\vec{x})
\right],
\end{eqnarray}
where the anti-symmetrization operator ${\cal{A}}$ can be written as
${\cal{A}}=1-P_{12}-P_{34}+P_{12}P_{34}$
for the $(N_{\uparrow},N_{\downarrow})=(2,2)$ system.
In Eq.~(\ref{eq_basis}), the $c_k$ denote expansion coefficients.
We parametrize the two-body potential $V_{\mathrm{tb}}$ by
a spherically symmetric 
attractive  Gaussian with depth $V_0$ ($V_0>0$) and range $r_0$,
\begin{eqnarray}
\label{eq_gauss}
V_{\mathrm{G}}(\vec{r}) = -V_0 \exp \left[ -
\left( \frac{r}{\sqrt{2}r_0} \right)^2 \right];
\end{eqnarray}
this interaction potential is convenient since the 
matrix elements $\langle \varphi_j | V_{\mathrm{G}} | \varphi_k \rangle$
are---for the $\varphi_k$
employed in this work---known 
analytically~\cite{cgbook} (see below for the definition
of the $\varphi_k$).
For a given range $r_0$, 
we adjust the depth $V_0$ such that $V_{\mathrm{G}}$ reproduces the
desired $s$-wave scattering length $a^{(aa)}$. For negative
(positive) $a^{(aa)}$,
we restrict ourselves to parameter combinations for which
$V_{\mathrm{G}}(\vec{r})$ supports no (one) $s$-wave free-space bound state.
For each $a^{(aa)}$, we consider a number of different
ranges $r_0$, $r_0 \ll a_{\mathrm{ho}}$, 
and extrapolate to the $r_0 \rightarrow 0$ limit
(see Sec.~\ref{sec_resultsn4} for details).

We employ two different classes
of non-orthogonal basis functions $\varphi_k$. 
For both classes, the Hamiltonian matrix elements and 
overlap matrix elements are known
analytically~\cite{cgbook}, 
thus reducing the problem of finding the eigenenergies 
and eigenfunctions to diagonalizing a generalized eigenvalue problem.
The first class of basis functions
$\varphi_k(\vec{x})$ has well defined 
angular momentum $L_{\mathrm{rel}}$ and natural parity,
while the latter class has neither well defined
angular momentum $L_{\mathrm{rel}}$ nor
well defined parity $\Pi_{\mathrm{rel}}$.

To describe natural parity states
with well defined 
angular momentum $L_{\mathrm{rel}}$ and
corresponding
projection quantum number $M_L$, we employ
the following basis functions~\cite{cgbook},
\begin{eqnarray}
\label{eq_cgbasis}
\varphi_k(\vec{x}) = 
|\vec{v}^{(k)}|^{L_{\mathrm{rel}}} {Y}_{L_{\mathrm{rel}}M_L}
%(\hat{\vec{v}}^{(k)}) 
(\hat{v}^{(k)}) 
\exp \left(-\frac{1}{2} \vec{x}^T \underline{A}^{(k)} \vec{x} \right),
\end{eqnarray}
where 
\begin{eqnarray}
\vec{v}^{(k)} = \sum_{j=1}^{N-1} u_j^{(k)} \vec{\rho}_j.
\end{eqnarray}
Here, the $u_j^{(k)}$, $j=1,\cdots,N-1$, 
define a $(N-1)$-dimensional parameter vector
that
determines how the angular momentum $L_{\mathrm{rel}}$ 
is distributed among the $(N-1)$
Jacobi vectors $\vec{\rho}_j$.
In Eq.~(\ref{eq_cgbasis}), $\underline{A}^{(k)}$ denotes a
$(N-1)\times(N-1)$-dimensional 
symmetric matrix, which is described
by
$N(N-1)/2$ independent parameters.
To get a physical interpretation of these parameters,
we rewrite the exponent on the right
hand side of Eq.~(\ref{eq_cgbasis})
in terms of a sum over
the square of interparticle distances $r_{ij}$ and 
$N(N-1)/2$ widths $d_{ij}^{(k)}$,
\begin{eqnarray}
\label{eq_exponent}
\frac{1}{2} \vec{x}^T \underline{A}^{(k)} \vec{x} = 
\frac{1}{2}
\sum_{i=1}^{N-1} \sum_{j=1}^{N-1} A_{ij}^{(k)} \vec{\rho}_i \cdot 
\vec{\rho}_j =
\nonumber \\
\sum_{i<j}^N \left(\frac{r_{ij}}{\sqrt{2}d_{ij}^{(k)}} \right)^2.
\end{eqnarray}
The explicit relationship
between the parameter matrix $\underline{A}^{(k)}$
and the widths $d_{ij}^{(k)}$ ($i<j$)
can be determined by expressing the
interparticle distance vectors $\vec{r}_{ij}$
in terms of the Jacobi vectors $\vec{x}$~\cite{cgbook}.
Equation~(\ref{eq_exponent}) illustrates that the $d_{ij}^{(k)}$
determine the widths of Gaussian functions in the interparticle
distance coordinates. 
In our calculations, we
choose a set of widths $d_{ij}^{(k)}$ for each basis function and construct
the matrix
 $\underline{A}^{(k)}$ from these.
The widths $d_{ij}^{(k)}$ themselves are---guided by physical 
arguments---determined semi-stochastically
following the schemes discussed in 
Refs.~\cite{cgbook,stec07c,stec08}.
For the $(N_{\uparrow},N_{\downarrow})=(2,2)$ 
system with small $r_0$ and small positive
$a^{(aa)}$, e.g.,
three-body and four-body bound states are 
absent~\cite{petr03,petr04aa}, implying that 
at most two of the widths $d_{13}^{(k)}$, $d_{14}^{(k)}$, $d_{23}^{(k)}$ and
$d_{24}^{(k)}$ 
(but not $d_{13}^{(k)}$ and $d_{14}^{(k)}$ simultaneously
or $d_{23}^{(k)}$ and $d_{24}^{(k)}$ simultaneously)
should be of the order of the two-body range
$r_0$  for a given $k$.
We use the basis functions given in Eq.~(\ref{eq_cgbasis})
to determine the eigenenergies of states 
with vanishing and finite $L_{\mathrm{rel}}$ and natural
parity, i.e., $\Pi_{\mathrm{rel}}=(-1)^{L_{\mathrm{rel}}}$.

To describe states with unnatural parity, we employ
basis functions $\varphi_k$ that are neither eigenfunctions
of the angular momentum operator $L_{\mathrm{rel}}$ nor the parity operator
$\Pi_{\mathrm{rel}}$~\cite{cgbook},
\begin{eqnarray}
\label{eq_cgbasis2}
\varphi_k(\vec{x})=
\exp \left(-\frac{1}{2}\vec{x}^T \underline{A}^{(k)} \vec{x} + 
(\vec{s}^{(k)})^T \vec{x} \right).
\end{eqnarray}
Here,
the quantity $\vec{s}^{(k)}$ consists of $N-1$ three-dimensional parameter
vectors and 
$(\vec{s}^{(k)})^T \vec{x}$ is just the dot product
between two $3(N-1)$ dimensional vectors.
The $3(N-1)$ parameters of $\vec{s}^{(k)}$
are, together with the $N(N-1)/2$ parameters
of the matrix $\underline{A}^{(k)}$,
optimized semi-stochastically~\cite{cgbook}.
Since the basis functions defined in
Eq.~(\ref{eq_cgbasis2}) are neither eigenfunctions of $L_{\mathrm{rel}}$,
$M_L$ nor $\Pi_{\mathrm{rel}}$, their use
allows for the determination of
the entire energy spectrum at once.
In Sec.~\ref{sec_resultsn4},
we employ the basis functions given in Eq.~(\ref{eq_cgbasis2})
to determine the energetically
lowest lying unnatural parity state of the four-fermion system
with negative $a^{(aa)}$.

Following the schemes outlined,
the determination of the four-fermion
energies corresponding to unnatural
parity states is significantly less numerically efficient 
than that of natural parity states.
This is, of course, not sursprising since only a ``fraction''
of the basis functions given in
Eq.~(\ref{eq_cgbasis2}) contributes to describing states with the
desired angular momentum, projection quantum number and parity.

\section{Results}
\label{sec_results}
This section summarizes the energetics of the three- and four-particle
equal-mass Fermi gas.

\subsection{Three-fermion system}
\label{sec_resultsn3} 
\begin{figure}
\vspace*{+.4cm}
\includegraphics[angle=0,width=70mm]{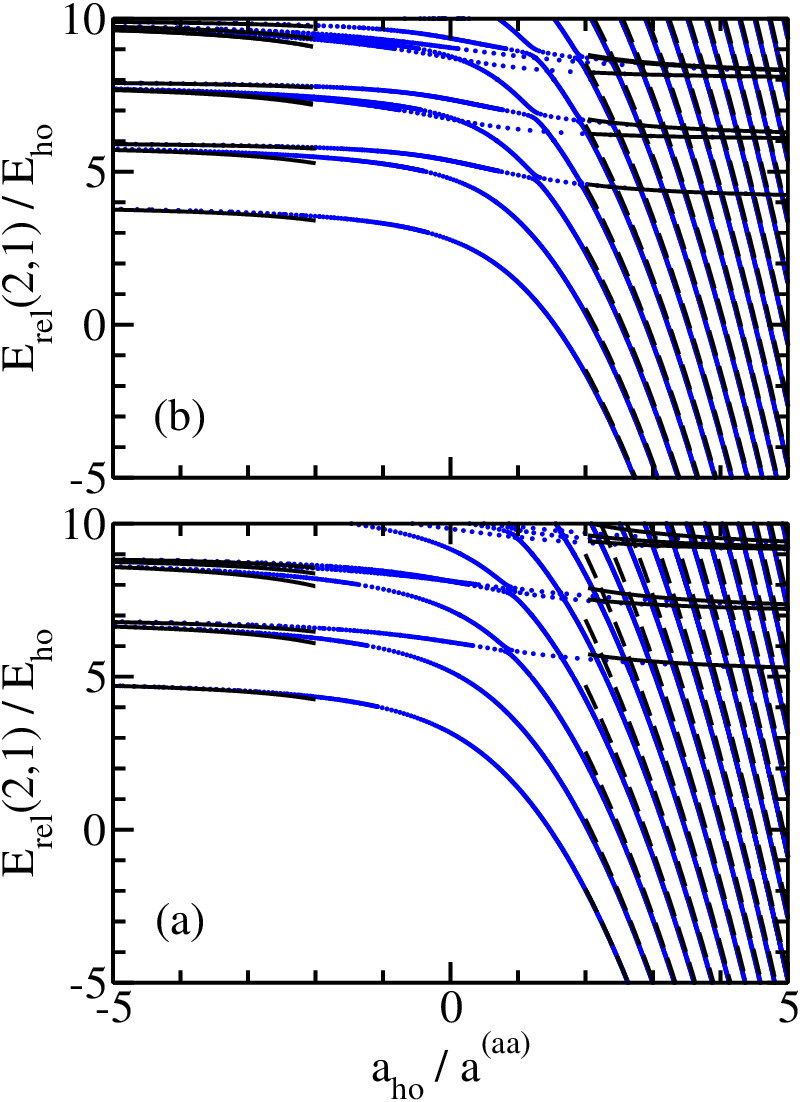}
\vspace*{0.1cm}
\caption{(Color online)
Three-fermion energies for (a) $(L_{\mathrm{rel}},\Pi_{\mathrm{rel}})=(0,+1)$ and
(b) $(L_{\mathrm{rel}},\Pi_{\mathrm{rel}})=(1,-1)$
as a function of the inverse $s$-wave scattering length $a_{\mathrm{ho}}/a^{(aa)}$.
Symbols show the essentially exact zero-range energies 
obtained by solving Eq.~(\ref{eq_kestner}). Solid lines
show the energies obtained by treating the non-interacting 
atomic Fermi gas
perturbatively for negative and positive $a^{(aa)}$
while dashed lines show the energies obtained by treating the 
effective atom plus dimer system perturbatively.
}\label{fig_energyn3}
\end{figure}

Figures~\ref{fig_energyn3}(a) and (b) show the eigenenergies $E_{\mathrm{rel}}(2,1)$
for states with
natural parity and $L_{\mathrm{rel}}=0$ and $1$
as a function of the inverse scattering length $1/a^{(aa)}$.
The symbols show the solutions to the coupled equations,
Eq.~(\ref{eq_kestner}), while the solid and dashed 
lines are obtained from our 
perturbative treatments of the atomic Fermi gas and the
effective atom plus dimer model,
respectively. 
Note that the perturbative treatment of the atomic Fermi gas (solid lines)
describes the energy levels corresponding to gas-like states
for negative as well as for positive $a^{(aa)}$ ($|a^{(aa)}|$ small).
In the following, we highlight selected characteristics of the
three-fermion
energy spectrum.

We first consider the weakly-interacting attractive Fermi gas.
In the non-interacting limit, $a^{(aa)} \rightarrow 0^-$,
the ground state has an energy of $E_{\mathrm{ni,rel}}=4\hbar \omega$
and is characterized by
$L_{\mathrm{rel}}=1$ and $\Pi_{\mathrm{rel}}=-1$ 
(see Fig.~\ref{fig_energyn3} and Table~\ref{tab_shift3}).
For the next family of
energies with $E_{\mathrm{ni,rel}}=5\hbar \omega$,
we have two natural parity states with $L_{\mathrm{rel}}=0$ 
and $L_{\mathrm{rel}}=2$, respectively,
and one unnatural parity state with $L_{\mathrm{rel}}=1$. 
The fact that the lowest non-interacting
$L_{\mathrm{rel}}=0$ state has a higher
energy than the lowest non-interacting
$L_{\mathrm{rel}}=1$ state can be 
understood intuitively 
by realizing that the two like atoms cannot both occupy the lowest single 
particle state. Within the hyperspherical description, this implies
that the $L_{\mathrm{rel}}=0$ state with $\lambda=0$ and $q=0$
is symmetry-forbidden (see Sec.~\ref{sec_hyperspherical})
and that the first symmetry-allowed $L_{\mathrm{rel}}=0$ 
state, which has $\lambda=2$ and $q=0$, lies $2\hbar \omega$ higher
in energy than the symmetry-forbidden $L_{\mathrm{rel}}=0$ 
state.

\begin{figure}
\vspace*{+.4cm}
\includegraphics[angle=0,width=70mm]{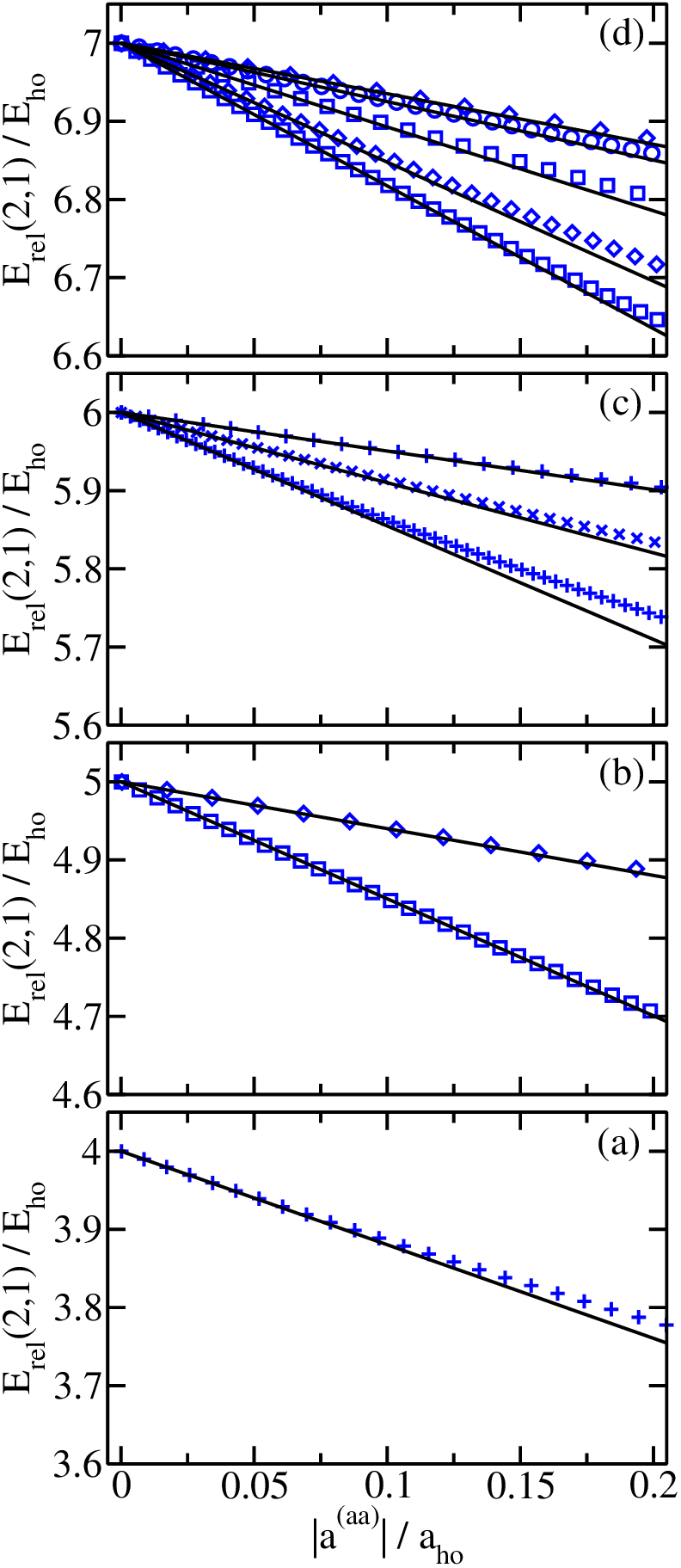}
\vspace*{0.1cm}
\caption{(Color online)
Three-fermion energies 
as a function of the absolute value of the $s$-wave scattering length
$|a^{(aa)}|$ for small $|a^{(aa)}|$, $a^{(aa)} \le 0$,
and 
(a) $E_{\mathrm{rel}}(2,1) \approx 4 \hbar \omega$,
(b) $E_{\mathrm{rel}}(2,1) \approx 5 \hbar \omega$,
(c) $E_{\mathrm{rel}}(2,1) \approx 6 \hbar \omega$, and
(d) $E_{\mathrm{rel}}(2,1) \approx 7 \hbar \omega$.
Squares, pluses, diamonds, crosses and 
circles show the essentially exact zero-range energies 
obtained by solving Eq.~(\ref{eq_kestner}) 
for $L_{\mathrm{rel}}=0-4$.
For comparison,
solid
lines
show the perturbative results for natural 
parity states.
The solid lines correspond
to (a) $(L_{\mathrm{rel}},\Pi_{\mathrm{rel}})=(1,-1)$; 
and from bottom to top to
(b) $(L_{\mathrm{rel}},\Pi_{\mathrm{rel}})=(0,+1)$ and
$(2,+1)$; 
(c) $(L_{\mathrm{rel}},\Pi_{\mathrm{rel}})=(1,-1)$,
$(3,-1)$, and
$(1,-1)$; and 
(d) $(L_{\mathrm{rel}},\Pi_{\mathrm{rel}})=(0,+1)$,
$(2,+1)$,
$(0,+1)$,
$(4,+1)$, and
$(2,+1)$.
}\label{fig_energyn3shift}
\end{figure}

The coefficients $c^{(1)}$ 
that determine the perturbative energy shifts $E^{(1)}$
of the atomic $(N_{\uparrow},N_{\downarrow})=(2,1)$
system are calculated semi-analytically 
following the scheme outlined 
in Secs.~\ref{sec_hamiltonian}
and \ref{sec_hyperspherical}, and
reported
in the last column of Table~\ref{tab_shift3}
for the first three energy families.
As already
mentioned, the unnatural 
parity
states  
of the three-fermion system are uneffected by the 
interactions~\cite{wern06,kest07,liu09}.
This behavior is specific to zero-range $s$-wave interactions
since a finite-range potential allows,
in general, for an energy shift due
to
$p$-wave or other higher partial wave interactions.
To illustrate the validity regime of the perturbative expressions,
Figs.~\ref{fig_energyn3shift}(a)-(d) 
show the small $|a^{(aa)}|$ region,
$a^{(aa)} \le 0$, of the
three-body energy spectrum as a function of $|a^{(aa)}|/a_{\mathrm{ho}}$ 
for the energies around the first four energy families
with $E_{\mathrm{rel}}(2,1) \approx 4 \hbar \omega$ to 
$E_{\mathrm{rel}}(2,1) \approx 7 \hbar \omega$.
As in Fig.~\ref{fig_energyn3}, the exact energies are shown by symbols
while the perturbative energies corresponding to  natural 
parity states
are shown by solid 
lines.
As expected, the perturbative treatment reproduces the
exact energies extremely well for small $|a^{(aa)}|/a_{\mathrm{ho}}$
and provides a semi-quantitatively correct description
up to 
$|a^{(aa)}| \approx 0.5 a_{\mathrm{ho}}$ (see also Fig.~\ref{fig_energyn3},
which shows the perturbative energies up to 
$a_{\mathrm{ho}}/|a^{(aa)}|=2$ or $|a^{(aa)}| = 0.5 a_{\mathrm{ho}}$).

\begin{figure}
\vspace*{.4cm}
\includegraphics[angle=0,width=70mm]{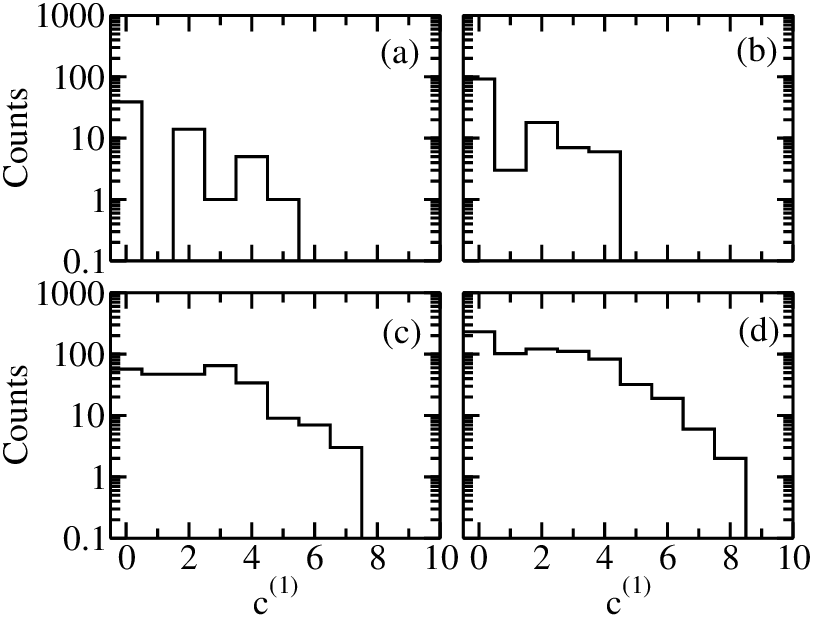}
\vspace*{0.1cm}
\caption{
Frequency of the coefficients $c^{(1)}$, which 
characterize the energy shifts $E^{(1)}$ of the weakly-interacting
atomic Fermi gas,
for the three- and four-fermion systems.
Panels~(a) and (b) show the distribution of the $c^{(1)}$ coefficients
for $E_{\mathrm{rel}}(2,1)\approx 7 \hbar \omega$ (fourth energy manifold
of the three-fermion system) and
for $E_{\mathrm{rel}}(2,1)\approx 8 \hbar \omega$ (fifth energy manifold
of the three-fermion system).
Panels~(c) and (d) show the distribution of the $c^{(1)}$ coefficients
for $E_{\mathrm{rel}}(2,2)\approx 19 \hbar \omega/2$ (fourth energy manifold
of the four-fermion system) and
for $E_{\mathrm{rel}}(2,2)\approx 21 \hbar \omega/2$ (fifth energy manifold
of the four-fermion system).
Note the log scale of the vertical axis.
}\label{fig_shiftn3}
\end{figure}

Table~\ref{tab_shift3} shows that the coefficients $c^{(1)}$
cover a wide range of values.
Within each energy family, 
the state shifted most strongly by the interactions is 
a natural parity state with the smallest allowed angular momentum
$L_{\mathrm{rel}}$.
To illustrate the increasing density of states and the spread of the
energy levels around the non-interacting degenerate energy manifold,
Figs.~\ref{fig_shiftn3}(a) and (b) show the frequency with which the
coefficients $c^{(1)}$ occur for
$E_{\mathrm{rel}}(2,1) \approx 7 \hbar \omega$ (fourth energy manifold)
and $E_{\mathrm{rel}}(2,1) \approx 8 \hbar \omega$ (fifth energy manifold),
respectively. In making this plot,
the $2L_{\mathrm{rel}}+1$ degeneracy of the energy levels has been taken into
account.
Since the unnatural parity states are not affected by the
zero-range interactions, the distribution of the $c^{(1)}$ coefficients 
shows a large amplitude for the $c^{(1)}=0$ bin.
Figures~\ref{fig_shiftn3}(a) and (b) show
that the spread of the coefficients
$c^{(1)}$ increases slightly as the energy manifold increases.
The primary characteristic of the distributions of the $c^{(1)}$
coefficients is, however, that the amplitude 
increases with increasing energy.

We now consider the weakly-repulsive regime, i.e., the regime where
$a^{(aa)}/a_{\mathrm{ho}} \ll1$ and $a^{(aa)}>0$.
In this regime, Fig.~\ref{fig_energyn3} shows two families of energy 
levels: (i)
energy levels with positive energy and
(ii) energy levels with negative energy.
The positive energy branches correspond
to states that describe
a gas of atoms; we refer to these states as 
the 
%``three fermionic atoms'' 
``gas-like state''
family.
The energies of this family are, in the
$a^{(aa)} \rightarrow 0^+$ limit, well described by treating the
atomic Fermi gas perturbatively (see solid lines in Fig.~\ref{fig_energyn3}).
The negative energy branches 
correspond to states that can be thought of
as consisting of a bound diatomic molecule and a
spare atom;
we refer to these states as 
``dimer plus atom'' family. 
In agreement with the literature (see, e.g., Ref.~\cite{petr03}),
Fig.~\ref{fig_energyn3} shows
that the formation of bound triatomic molecules is prohibited by the
Pauli exclusion principle or the so-called Pauli pressure.
For small and positive $a^{(aa)}$,
the perturbative energy shifts for the energy levels
with $L_{\mathrm{rel}}=0$ and
$\Pi_{\mathrm{rel}}=+1$ [dashed lines in Fig.~\ref{fig_energyn3}(a)]
are
calculated using Eq.~(\ref{eq_shiftdimer}) with $k=ad$.
The perturbative approach, applied to the effective
model Hamiltonian $H^{\mathrm{eff}}$,
predicts no energy shift for
states with $L_{\mathrm{rel}}>0$.
This follows directly from the fact that we parametrized the
effective atom-dimer interaction through a zero-range 
$s$-wave potential.
Thus, the ``bending'' of the
dashed lines in Fig.~\ref{fig_energyn3}(b) for $L_{\mathrm{rel}}=1$
(and in general, $L_{\mathrm{rel}}>0$)
and positive $a^{(aa)}$ is solely due to the internal 
energy of the dimer and not due to the effective atom-dimer interaction.
We find that the perturbative treatment
provides a qualitatively correct 
description up to $a^{(aa)} \approx 0.5 a_{\mathrm{ho}}$.

The effective model Hamiltonian $H^{\mathrm{eff}}$
also provides an intuitive
picture for why the lowest $L_{\mathrm{rel}}=0$ state has a lower energy 
than the lowest $L_{\mathrm{rel}}=1$ state as $a^{(aa)}\rightarrow 0^+$. 
In the $a^{(aa)}\rightarrow 0^+$
limit, the diatomic molecule has vanishing angular
momentum and the angular momentum $L_{\mathrm{rel}}$
must be carried by the atom-dimer distance
vector.
Thus, the energy of the lowest state with $L_{\mathrm{rel}}=1$
lies approximately
$\hbar \omega$ above the lowest state with $L_{\mathrm{rel}}=0$,
the energy of the lowest state with $L_{\mathrm{rel}}=2$
lies approximately 
$\hbar \omega$ above the lowest state with $L_{\mathrm{rel}}=1$,
and so on.
The parity inversion of the energetically lowest lying state
($L_{\mathrm{rel}}=1$ and $\Pi_{\mathrm{rel}}=-1$ in the $a^{(aa)} \rightarrow 0^-$ limit,
and $L_{\mathrm{rel}}=0$ and $\Pi_{\mathrm{rel}}=+1$ in the $a^{(aa)} \rightarrow 0^+$ limit)
occurs at $a_{\mathrm{ho}}/a^{(aa)} \approx 1$
and has already been pointed out by a number of 
works~\cite{kest07,stet07,stec08}.

\begin{figure}
\vspace*{+.4cm}
\includegraphics[angle=0,width=70mm]{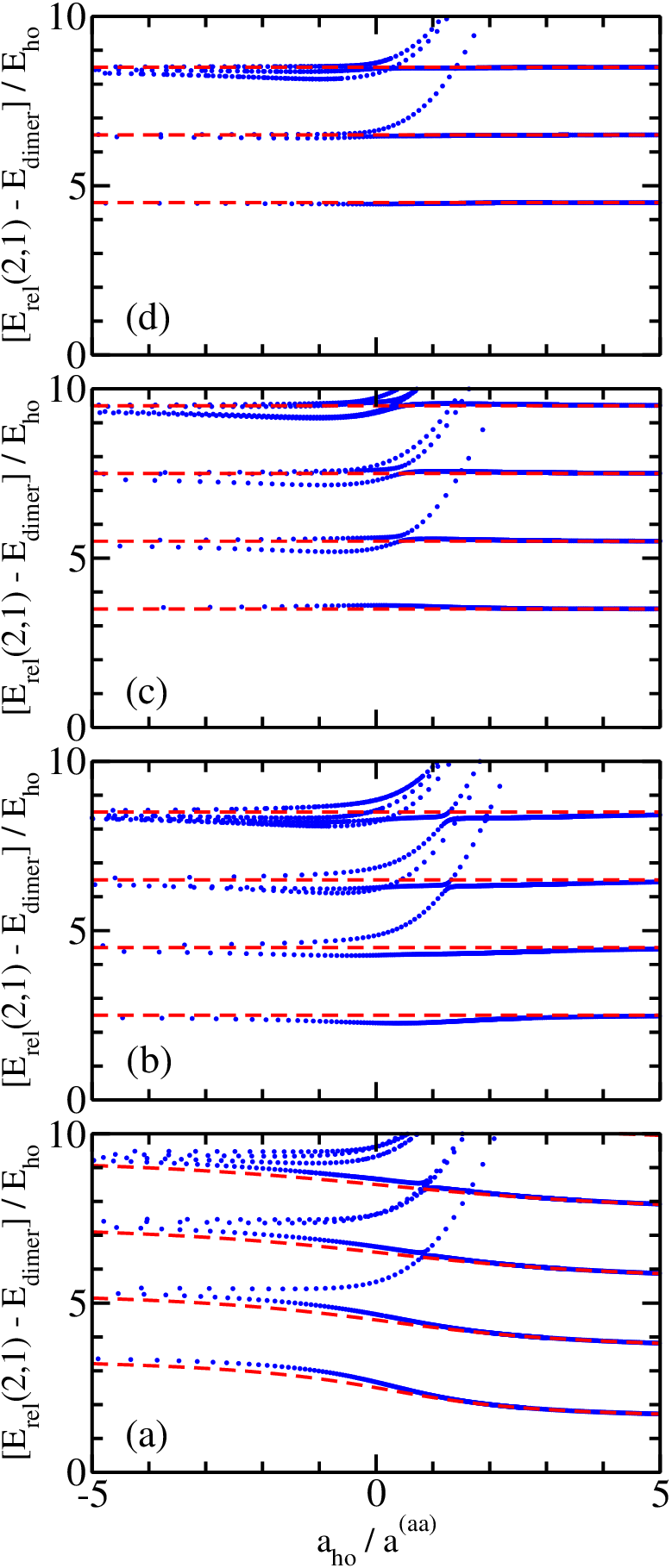}
\vspace*{0.1cm}
\caption{(Color online)
Symbols show the scaled energies $E_{\mathrm{rel}}(2,1)-E_{\mathrm{dimer}}$ 
as a function of the inverse scattering length $a_{\mathrm{ho}}/a^{(aa)}$
for 
(a) $(L_{\mathrm{rel}},\Pi_{\mathrm{rel}})=(0,+1)$,
(b) $(L_{\mathrm{rel}},\Pi_{\mathrm{rel}})=(1,-1)$,
(c) $(L_{\mathrm{rel}},\Pi_{\mathrm{rel}})=(2,+1)$, and
(d) $(L_{\mathrm{rel}},\Pi_{\mathrm{rel}})=(3,-1)$.
For comparison,
dashed lines show the energies predicted by
the effective atom-dimer model, which provides
a semi-quantitative description of the states belonging
to the ``atom plus dimer'' family (see text for details). 
}\label{fig_energyn3scaled}
\end{figure}

Motivated by the fact that 
the energy spectrum 
of the three-fermion
system can be described by an effective atom plus dimer model
in the $a^{(aa)}\rightarrow 0^+$
limit, 
symbols in Figs.~\ref{fig_energyn3scaled}(a)-(d)
show the quantity $E_{\mathrm{rel}}(2,1)-E_{\mathrm{dimer}}$
as a function of the inverse atom-atom
scattering length $1/a^{(aa)}$ for $L_{\mathrm{rel}}=0-3$.
Here, $E_{\mathrm{dimer}}$ denotes the lowest eigenenergy of the trapped
$s$-wave interacting atom-atom system, i.e., the lowest
eigenenergy of Eq.~(\ref{eq_busch}) with $k=aa$.
The quantity $E_{\mathrm{rel}}(2,1)-E_{\mathrm{dimer}}$ has been investigated
previously~\cite{kest07,stec08} 
and has been termed the universal energy crossover curve
in Ref.~\cite{stec08}.
Figures~\ref{fig_energyn3scaled}(b)-(d) show the existence of
a family of states for $L_{\mathrm{rel}}>0$ whose scaled energies
are nearly independent of the $s$-wave scattering length $a^{(aa)}$.
These scaled energies are approximately given by 
$(2n_{\mathrm{eff}} +L_{\mathrm{rel}}+3/2) \hbar \omega$
[see dashed lines in Figs.~\ref{fig_energyn3scaled}(b)-(d)].
The fairly good agreement between the symbols and the dashed lines 
reflects the fact that a subset of the three-fermion energies
can be described to a fairly good
approximation
by treating the three-fermion system as consisting of
a bound trapped $s$-wave dimer plus a 
non-interacting spare atom.
Figures~\ref{fig_energyn3scaled}(b)-(d) show
that this effective dimer-atom
description improves with increasing 
$L_{\mathrm{rel}}$. 

The fact that a subset of states 
of the three-particle spectrum is 
reasonably well described by the effective
dimer plus atom model suggests that the three-particle energy spectrum
can be described in terms of avoided crossings between ``atom 
plus dimer'' states
and 
%``three fermionic atoms'' 
``gas-like''
states.
An interpretation along this line has been
quantified by von Stecher~\cite{stec08d} 
who applied a diabatization scheme.
Here, we do not follow the diabatization scheme but
instead
offer a qualitative discussion 
of the natural parity three-fermion spectrum
with $L_{\mathrm{rel}}>0$. We make four observations:
(i)
A sequence of states 
has an energy $E_{\mathrm{rel}}(2,1)$ of approximately
$(2+L_{\mathrm{rel}}+2n)\hbar \omega$ at unitarity,
an energy $E_{\mathrm{rel}}(2,1)$ of approximately
$(3+L_{\mathrm{rel}}+2n)\hbar \omega$
in the $a^{(aa)} \rightarrow 0^-$ limit,
and an energy $E_{\mathrm{rel}}(2,1)$ of approximately 
$E_{\mathrm{dimer}}+(3/2+L_{\mathrm{rel}}+2n)\hbar \omega$
in the $a^{(aa)} \rightarrow 0^+$ limit
[see Fig.~\ref{fig_energyn3}(b); the states
discussed here are those with approximately
constant $E_{\mathrm{rel}}(2,1)-E_{\mathrm{dimer}}$, see 
Figs.~\ref{fig_energyn3scaled}(b)-(d)].
(ii)
For each $L_{\mathrm{rel}}$,
the degeneracy of the energy families with $E_{\mathrm{ni,rel}}=
(3+L_{\mathrm{rel}}+2n)\hbar \omega$, $n=0,1,\cdots$,
increases by one (or $2L_{\mathrm{rel}}+1$ if
the degeneracy of the different $M_L$ values
is accounted for
explicitly) as $n$ increases by one (see Fig.~\ref{fig_energyn3},
Fig.~\ref{fig_energyn3scaled} and Table~\ref{tab_shift3}).
(iii)
The energy spectrum corresponding to
%``three fermionic atoms'' 
``gas-like'' states
has to be identical in the limits
$a^{(aa)} \rightarrow 0^-$ 
and
$a^{(aa)} \rightarrow 0^+$.
(iv) It can be easily checked that (i)-(iii)
are consistent with the fact that
the energy $E_{\mathrm{rel}}(2,1)$ of all but one level of each non-interacting 
manifold with a given $L_{\mathrm{rel}}$
decreases
by $\hbar \omega$ when going from $a^{(aa)} \rightarrow 0^-$
to $a^{(aa)} \rightarrow \infty$
and by another $\hbar \omega$ 
when going from $a^{(aa)} \rightarrow \infty$
to $a^{(aa)} \rightarrow 0^+$.
The dropping of the energies by $2 \hbar \omega$
as $a^{(aa)}$ changes from $0^-$ through $\pm \infty$ to $0^+$
is similar to the $2\hbar \omega$ dropping of the excited state
$s$-wave energies
of the two-particle system 
(see dashed lines in Fig.~\ref{fig_energyn2})
and can be interpreted as one 
%spin-up/spin-down 
pair (consisting of a spin-up atom and a spin-down atom)
feeling the $s$-wave interaction 
while the other spin-up
atom
%/spin-down pair 
carrying
the angular momentum.

\begin{figure}
\vspace*{+.4cm}
\includegraphics[angle=0,width=70mm]{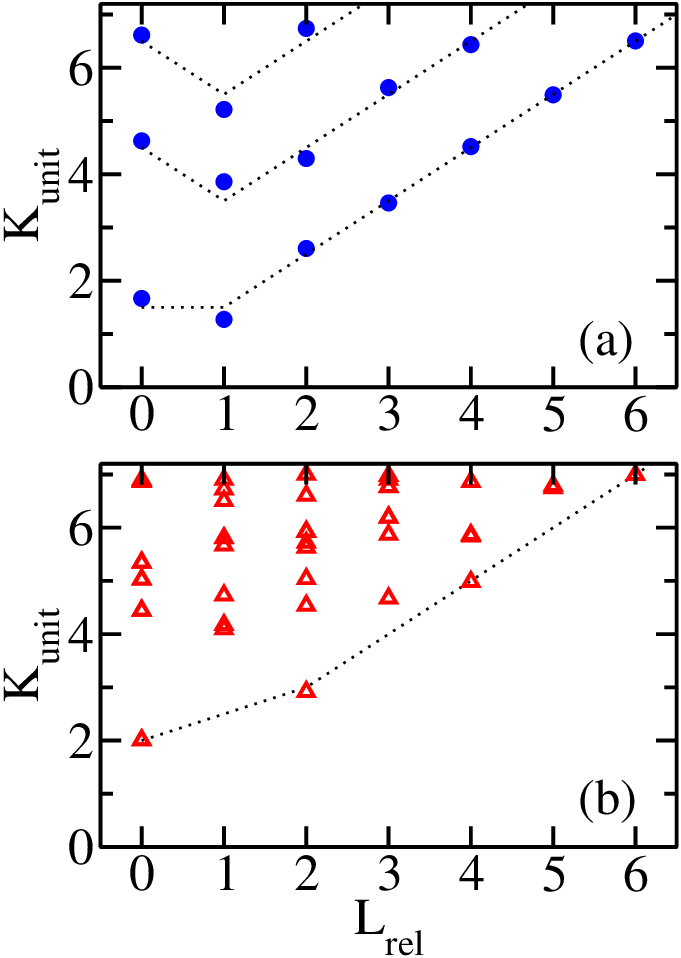}
\vspace*{0.1cm}
\caption{(Color online)
Symbols show the coefficients $K_{\mathrm{unit}}$ that correspond to natural
parity
states for (a) the three-fermion system 
and (b) the four-fermion system
as a function of the angular momentum quantum number $L_{\mathrm{rel}}$.
Dotted lines show the coefficients $K_{\mathrm{unit,model}}$
(see text for details).
}\label{fig_energyunit}
\end{figure}

Interestingly, the observations described in the previous
paragraph allow for an approximate determination
of the $K_{\mathrm{unit}}$ coefficients [see
Eq.~(\ref{eq_energyunit})].
Observation (i) 
implies that
the lowest $K_{\mathrm{unit}}$
coefficient for a given $L_{\mathrm{rel}}$ is approximately
given by $K_{\mathrm{unit,model}}=L_{\mathrm{rel}}+1/2$
in the large $L_{\mathrm{rel}}$ limit. 
As discussed in Sec.~\ref{sec_hyperspherical}, 
each $K_{\mathrm{unit}}$ coefficient determines the starting point
of a ladder of energy levels, which
are spaced by $2\hbar \omega$
and which 
are associated with an increasing number of nodes along the
hyperradial coordinate $R$.
It is evident from Figs.~\ref{fig_energyn3scaled}(b)-(d)
that these states, which are characterized by the same hyperangular
quantum number but different hyperradial quantum numbers $q$, transform into 
atom plus dimer states in the $a^{(aa)} \rightarrow 0^+$ limit,
which are characterized by the effective orbital angular
momentum quantum number $l_{\mathrm{eff}}=0$
and different
radial quantum numbers $n_{\mathrm{eff}}$.  
Interpreting the atom-dimer distance coordinate as the 
hyperradial coordinate in the $a^{(aa)} \rightarrow 0^+$
limit, the identification $q \leftrightarrow n_{\mathrm{eff}}$ suggests
itself.
Using observation (iv),
the remaining $K_{\mathrm{unit}}$
coefficients at unitarity 
are approximately given by
$K_{\mathrm{unit,model}}=L_{\mathrm{rel}}+1/2+2n$ for $L_{\mathrm{rel}}>0$
and natural parity states [for each $n=1,2,\cdots$,
the $q$ quantum number
in Eq.~(\ref{eq_energyunit}) takes the values $q=0,1,\cdots$].
Figure~\ref{fig_energyunit}(a) shows 
that the difference between $K_{\mathrm{unit}}$ 
(symbols) and $K_{\mathrm{unit,model}}$ (dotted lines)
decreases as $L_{\mathrm{rel}}$ increases.
The difference between $K_{\mathrm{unit}}$ and $K_{\mathrm{unit,model}}$
has previously
been quantified by Werner and Castin within a semi-classical WKB
framework~\cite{wern06a}.

The $L_{\mathrm{rel}}=0$ spectrum is different from the 
$L_{\mathrm{rel}}>0$ spectra for two reasons. 
First, the effective atom-dimer
system is described by an effective atom-dimer 
$s$-wave scattering length $a^{(ad)}$,
which leads to a decrease
of approximately 
$\hbar \omega$ of the energy levels belonging to the 
atom plus dimer family as $a^{(aa)}$ changes from $\infty$ to
$0^+$.
Second, the symmetry constraint in the $a^{(aa)} \rightarrow 0^-$
limit pushes the energy of the lowest $L_{\mathrm{rel}}=0$ state 
up by $\hbar \omega$ compared to that with $L_{\mathrm{rel}}=1$.
Dashed lines in Fig.~\ref{fig_energyn3scaled}(a) show the eigenenergies
of the effective atom plus dimer model, i.e., the
eigenenergies of Eq.~(\ref{eq_busch}) for $k=ad$ with $a^{(ad)}$
given by Eq.~(\ref{eq_atomdimer}).
It can be seen that the agreement between the energies of this
effective model and a subset of the full three-fermion
energies is good for small positive $a^{(aa)}$
and qualitatively correct throughout the entire crossover regime.
Using the effective atom plus dimer model, the energy
at unitarity of a subset of states is approximately given by
$E_{\mathrm{rel}}(2,1) \approx E_{\mathrm{dimer}}+(2q +5/2)\hbar \omega$, 
implying 
$K_{\mathrm{unit,model}}=3/2$. For comparison, the exact value is 
$K_{\mathrm{unit}}=1.666$~\cite{wern06a}.
The other $K_{\mathrm{unit}}$ coefficients for $L_{\mathrm{rel}}=0$ can be estimated
by using that 
the energy of one subset
of states drops by $\hbar \omega$ in going from $a^{(aa)} \rightarrow
\infty$ to $a^{(aa)} \rightarrow 0^+$,
implying
$E_{\mathrm{unit,rel}}(2,1) \approx (5/2+2q+2n)\hbar \omega$
with $n=1,2,\cdots$
and $q=0,1,\cdots$.
This estimate yields
$K_{\mathrm{unit,model}}=5/2+2n$ for 
$n=1,2,\cdots$. 
Figure~\ref{fig_energyunit}(a)
shows that the
$K_{\mathrm{unit,model}}$ (dotted lines) reproduce the exact
$K_{\mathrm{unit}}$ coefficients (symbols) fairly well.

\subsection{Four-fermion system}
\label{sec_resultsn4} 
This section discusses the 
energy spectrum of the 
four-fermion system throughout the BCS-BEC crossover.
We primarily focus on the energies corresponding to
natural parity states but also consider those corresponding to
unnatural states.
To determine the energy spectrum
corresponding to states with natural parity, we 
use the stochastic
variational approach with the basis functions given in
Eq.~(\ref{eq_cgbasis});
our basis set optimization
either treats one state at a time or a subset of states simultaneously.
For a given atom-atom scattering 
length $a^{(aa)}$,
we determine the energies
for various ranges $r_0$, $r_0=0.01a_{\mathrm{ho}}-0.09a_{\mathrm{ho}}$,
of the two-body Gaussian interaction potential
and extrapolate the finite-range energies to $r_0 \rightarrow 0$.
For negative scattering lengths $a^{(aa)}$, we find that the
four-fermion energies depend linearly on $r_0$ for all $L_{\mathrm{rel}}$
considered. In this regime, we typically calculate the energies for three
different $r_0$ and then determine the 
$r_0 \rightarrow 0$
energies by performing a linear fit.
For positive scattering lengths $a^{(aa)}$, the 
energetically low-lying part of the spectrum is dominated by the
``internal'' energy of the dimer(s) formed.
As discussed in more detail below,
the lowest energy family for 
even $L_{\mathrm{rel}}$ can be described by an effective two-boson model
while the lowest energy family for odd $L_{\mathrm{rel}}$
consists of
states that can be thought of as consisting of a dimer and two atoms
as $a^{(aa)}\rightarrow 0^+$ (see Sec.~\ref{sec_theory} and below).
Correspondingly, 
for positive $a^{(aa)}$ and even $L_{\mathrm{rel}}$,
we 
subtract twice the dimer binding energy from the four-fermion
energies 
for each $r_0$ and extrapolate the scaled
four-fermion energies to the $r_0 \rightarrow 0$ limit.
We typically consider five different $r_0$ and extract the
scaled zero-range energies by performing a 
quadratic fit to the scaled finite-range four-fermion energies.
The zero-range four-fermion energies 
themselves are then obtained by adding twice
the zero-range dimer energy, i.e., $2E_{\mathrm{dimer}}$.
For odd $L_{\mathrm{rel}}$, 
we subtract (and later add) the dimer energy
as opposed to twice the dimer energy but proceed analogously
otherwise.

\begin{figure}
\vspace*{+.4cm}
\includegraphics[angle=0,width=70mm]{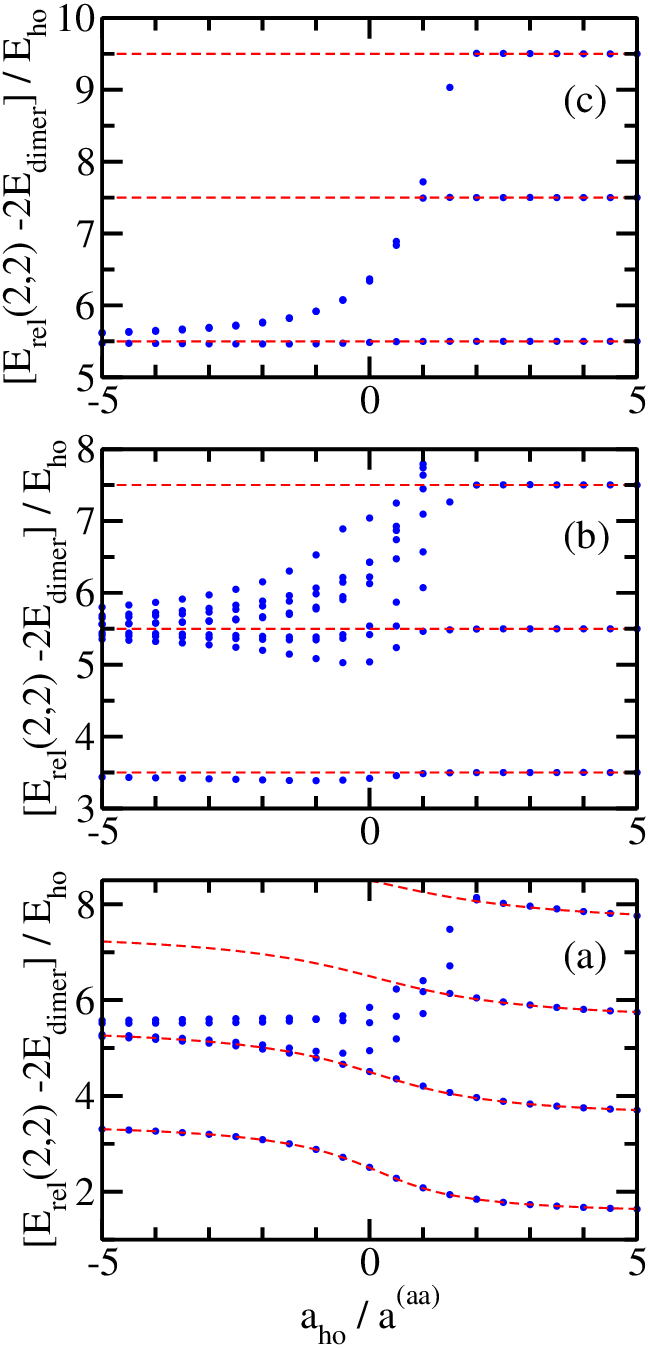}
\vspace*{0.1cm}
\caption{(Color online)
Scaled four-fermion energies $E_{\mathrm{rel}}(2,2)-2E_{\mathrm{dimer}}$
as a function of 
$a_{\mathrm{ho}}/a^{(aa)}$ for (a) $(L_{\mathrm{rel}},\Pi_{\mathrm{rel}})=(0,+1)$,
(b) $(L_{\mathrm{rel}},\Pi_{\mathrm{rel}})=(2,+1)$ and
(c) $(L_{\mathrm{rel}},\Pi_{\mathrm{rel}})=(4,+1)$.
Symbols show the numerically determined energies
while dashed lines show the energies predicted by the 
effective dimer-dimer model 
(see text for details).
}\label{fig_energyn4scale}
\end{figure}

The energies of a large number
of energy levels corresponding to natural parity
states are
reported in the auxiliary materials~\cite{epaps}
for
atom-atom scattering lengths $a^{(aa)}$
ranging from $0^-$ over $\infty$ to 
$0.2 a_{\mathrm{ho}}$ for $L_{\mathrm{rel}} \le 4$. To the best of our
knowledge, these are the first comprehensive
benchmark results for the four-fermion system
with finite angular momentum 
throughout the crossover region.
Tables~\ref{tab_energyunit}
\begin{table}
\caption{Extrapolated zero-range 
energies $E_{\mathrm{unit,rel}}(2,2)$
%for the four-fermion system  
for natural parity states with $L_{\mathrm{rel}} \le 4$
%at unitarity 
[$E_{\mathrm{unit,rel}}(2,2) \le 8.5 \hbar \omega$]. 
The uncertainty of the 
%extrapolated zero-range 
energies
is estimated to be in the last digit reported.
For comparison, the lowest energy 
of the $(L_{\mathrm{rel}},\Pi_{\mathrm{rel}})=(1,+1)$ state
%corresponding to
%the unnatural parity state with $L_{\mathrm{rel}}=1$ 
is $E_{\mathrm{unit,rel}}=5.088(20) \hbar \omega$.}
\begin{ruledtabular}
\begin{tabular}{lllll}
  $L_{\mathrm{rel}}$ & $\Pi_{\mathrm{rel}}$ &  $q$ & $K_{\mathrm{unit}}$ & 
$E_{\mathrm{unit,rel}}/E_{\mathrm{ho}}$  \\
\hline
0 & $+1$ &  0 & 2.009 & 3.509  \\ 
0 & $+1$ &  1 & 2.010 & 5.510  \\ 
0 & $+1$ &  0 & 4.444 & 5.944 \\ 
0 & $+1$ &  0 & 5.029 & 6.529  \\ 
0 & $+1$ &  0 & 5.347 & 6.847 \\ 
0 & $+1$ &  2 & 2.017 & 7.517 \\ 
0 & $+1$ &  1 & 4.446 & 7.946 \\ 
0 & $+1$ &  0 & 6.864 & 8.364 \\ 
0 & $+1$ &  0 & 6.905 & 8.405 \\ 
\hline
1 & $-1$ &  0 & 4.098 & 5.598  \\
1 & $-1$ &  0 & 4.176 & 5.676  \\
1 & $-1$ &  0 & 4.730 & 6.230  \\
1 & $-1$ &  0 & 5.669 & 7.169 \\ 
1 & $-1$ &  0 & 5.807 & 7.307 \\ 
1 & $-1$ &  1 & 4.101 & 7.601 \\ 
1 & $-1$ &  1 & 4.180 & 7.680 \\ 
1 & $-1$ &  0 & 6.505 & 8.005 \\ 
1 & $-1$ &  0 & 6.724 & 8.224 \\ 
1 & $-1$ &  1 & 4.732 & 8.232 \\ 
1 & $-1$ &  0 & 6.904 & 8.404 \\ 
\hline
2 & $+1$ &  0 & 2.918 & 4.418 \\ 
2 & $+1$ &  0 & 4.539 & 6.039\\ 
2 & $+1$ &  1 & 2.920 & 6.420\\ 
2 & $+1$ &  0 & 5.039 & 6.539\\ 
2 & $+1$ &  0 & 5.629 & 7.129\\ 
2 & $+1$ &  0 & 5.722 & 7.222\\ 
2 & $+1$ &  0 & 5.925 & 7.425\\ 
2 & $+1$ &  0 & 5.927 & 7.427\\ 
2 & $+1$ &  1 & 4.542 & 8.042 \\ 
2 & $+1$ &  0 & 6.707 & 8.207 \\ 
2 & $+1$ &  2 & 2.924 & 8.424 \\ 
2 & $+1$ &  0 & 7.001 & 8.501 \\ 
\hline
3 & $-1$ &  0 & 4.676 & 6.176  \\ 
3 & $-1$ &  0 & 5.871 & 7.371  \\ 
3 & $-1$ &  0 & 6.191 & 7.691 \\ 
3 & $-1$ &  0 & 6.194 & 7.694 \\ 
3 & $-1$ &  1 & 4.678 & 8.178 \\ 
3 & $-1$ &  0 & 6.764 & 8.264 \\ 
3 & $-1$ &  0 & 6.771 & 8.271 \\ 
3 & $-1$ &  0 & 6.904 & 8.404 \\ 
3 & $-1$ &  0 & 6.977 & 8.477 \\ 
\hline
4 & $+1$ &  0 & 4.985 & 6.485  \\ 
4 & $+1$ &  0 & 5.838 & 7.338 \\ 
4 & $+1$ &  0 & 5.868 & 7.368 \\
4 & $+1$ &  0 & 6.865 & 8.365 \\
4 & $+1$ &  1 & 4.984 & 8.484 \\
\end{tabular}
\end{ruledtabular}
\label{tab_energyunit}
\end{table}
and \ref{tab_energyunit2}
\begin{table}
\caption{
Lowest two extrapolated zero-range energies $E_{\mathrm{unit,rel}}(2,2)$
for the four-fermion system 
with natural parity and $L_{\mathrm{rel}}=5-8$
at unitarity. 
The uncertainty of the extrapolated zero-range energies
is estimated to be in the last or second last digit reported.
}
\begin{ruledtabular}
\begin{tabular}{lllll}
  $L_{\mathrm{rel}}$ & $\Pi_{\mathrm{rel}}$ &  $q$ & $K_{\mathrm{unit}}$ & 
$E_{\mathrm{unit,rel}}/E_{\mathrm{ho}}$  \\
\hline
\hline
5 & $-1$ &  0 & 6.745 & 8.245 \\
5 & $-1$ &  0 & 6.790 & 8.290 \\
\hline
6 & $+1$ &  0 & 6.996 & 8.496  \\ 
6 & $+1$ &  0 & 7.781 & 9.281  \\ 
\hline
7 & $-1$ &  0 & 8.769 & 10.269  \\ 
7 & $-1$ &  0 & 8.777 & 10.277  \\ 
\hline
8 & $+1$ &  0 & 8.998 & 10.498   \\
8 & $+1$ &  0 & 9.775 & 11.275  
\end{tabular}
\end{ruledtabular}
\label{tab_energyunit2}
\end{table}
summarize the energies of the four-fermion
system at unitarity for natural parity states
with $L_{\mathrm{rel}}=0$ to $L_{\mathrm{rel}}=8$ as well as for one unnatural parity state.
For $L_{\mathrm{rel}}>0$, these are the first
results at unitarity.
We estimate that our extrapolated zero-range energies 
for the energetically lowest-lying $L_{\mathrm{rel}}=0$ state 
is
accurate to better than 0.1\%
for most scattering lengths $a^{(aa)}$, including infinitely large 
$a^{(aa)}$. Near the avoided crossings
around $a^{(aa)} \approx a_{\mathrm{ho}}$ (see, e.g., 
Fig.~\ref{fig_energyn4scale}), however,
the accuracy decreases by up to
an order of magnitude.
Generally speaking, 
we find that the accuracy of the extrapolated zero-range 
energies also decreases for energetically higher lying states
and for states with larger $L_{\mathrm{rel}}$.
The eigenenergies reported 
in Tables~\ref{tab_energyunit} and \ref{tab_energyunit2}
are labeled by the
hyperradial quantum number $q$.
Following Eq.~(\ref{eq_energyunit}), we identify 
this quantum number by looking for 
$2 \hbar \omega$ spacings between energy pairs.
Inspection of Table~\ref{tab_energyunit} shows 
that the energies with $q>0$ lie,
within our numerical accuracy, $2 \hbar \omega$
above an energy with $q-1$ (see the nearly identical $K_{\mathrm{unit}}$
coefficients in the fourth column of 
Tables~\ref{tab_energyunit} and \ref{tab_energyunit2}).
Figure~\ref{fig_energyunit}(b) shows the $K_{\mathrm{unit}}$ coefficients
corresponding to natural parity states of the four-fermion system
as a function of $L_{\mathrm{rel}}$. Compared to the three-fermion system,
the 
four-fermion system exhibits a notably denser energy spectrum
at unitarity [see Fig.~\ref{fig_energyunit}(a) and \ref{fig_energyunit}(b)].

\begin{figure}
\vspace*{+.4cm}
\includegraphics[angle=0,width=70mm]{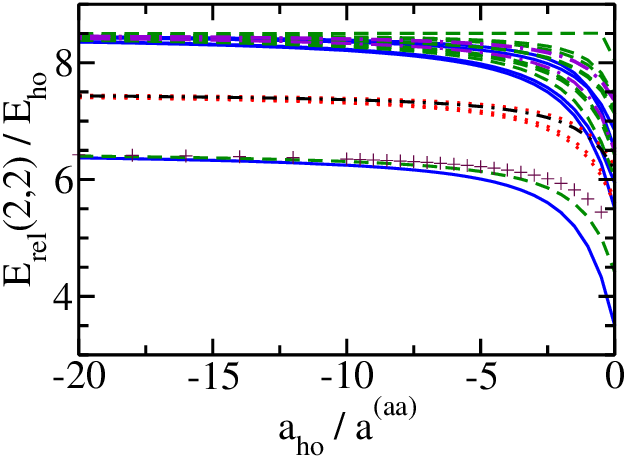}
\vspace*{0.1cm}
\caption{(Color online)
Four-fermion energies $E_{\mathrm{rel}}(2,2)$
as a function of the inverse $s$-wave scattering length 
$a_{\mathrm{ho}}/a^{(aa)}$ with $1/a^{(aa)} \le 0$.
For $a^{(aa)} \rightarrow 0^-$, all energy levels 
corresponding to 
natural parity states 
around $E_{\mathrm{ni,rel}}=13 \hbar \omega/2$,
$15 \hbar \omega/2$ and $17 \hbar \omega/2$ are shown.
Solid, dotted, dashed, dash-dotted and dash-dash-dotted lines show 
the zero-range
energies corresponding to natural parity states with $L_{\mathrm{rel}}=0$
to $L_{\mathrm{rel}}=4$.
Pluses show the energy of the energetically lowest-lying unnatural parity
state with $L_{\mathrm{rel}}=1$.
}\label{fig_energyn4}
\end{figure}

Lines
in Fig.~\ref{fig_energyn4} show the extrapolated
zero-range energies for the four-fermion system
corresponding to natural parity states with $L_{\mathrm{rel}}=0$ to $L_{\mathrm{rel}}=4$
as a function of the inverse $s$-wave scattering length $1/a^{(aa)}$
for negative $a^{(aa)}$.
In the $a^{(aa)} \rightarrow 0^-$ limit,
the three energetically lowest-lying
four-fermion energy manifolds 
around $E_{\mathrm{ni,rel}}(2,2)=13 \hbar \omega/2$,
$15 \hbar \omega/2$  and
$17 \hbar \omega/2$ 
consist of two, four and 15 states, respectively
(here, the $2L_{\mathrm{rel}}+1$ degeneracy due to the $M_L$ quantum number
is not included in counting the states;
see also Table~\ref{tab_shift4}).
For comparison, pluses in Fig.~\ref{fig_energyn4}
show the energetically lowest lying unnatural
parity state with ($L_{\mathrm{rel}},\Pi_{\mathrm{rel}})=(1,+1)$; in this
case, the energies 
are calculated for a 
Gaussian two-body potential with small but finite $r_0$
and have not been extrapolated to the
$r_0 \rightarrow 0$ limit.
Figure~\ref{fig_energyn4} shows 
that the three energy manifolds remain distinguishable
up to $a_{\mathrm{ho}}/a^{(aa)} \approx -2$ but start to
overlap notably in the strongly-interacting regime.

\begin{figure}
\vspace*{+1.4cm}
\includegraphics[angle=0,width=70mm]{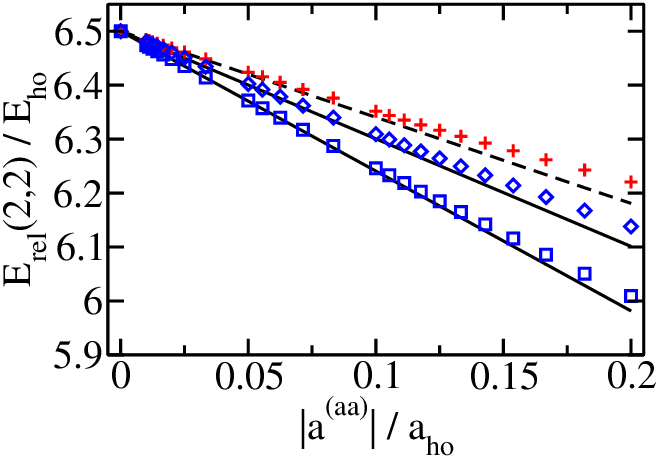}
\vspace*{0.1cm}
\caption{(Color online)
Four-fermion energies $E_{\mathrm{rel}}(2,2)$
as a function of the absolute value of the $s$-wave scattering length
$|a^{(aa)}|$ for small $|a^{(aa)}|$, $a^{(aa)} \le 0$,
and  $E_{\mathrm{rel}}(2,2) \approx 13 \hbar \omega/2$.
Squares, diamond and pluses 
show the numerically determined four-fermion energies 
for $(L_{\mathrm{rel}},\Pi_{\mathrm{rel}})=(0,+1)$, $(L_{\mathrm{rel}},\Pi_{\mathrm{rel}})=(2,+1)$ 
and $(L_{\mathrm{rel}},\Pi_{\mathrm{rel}})=(1,+1)$, respectively.
Solid and dashed
lines
show the perturbative results for natural 
and unnatural
parity states, respectively.
}\label{fig_energyn4shiftbcs}
\vspace*{0.4cm}
\end{figure}

We now discuss the weakly-attractive regime of the four-fermion
energy spectrum in more
detail.
The 
coefficients $c^{(1)}$ that determine the
perturbative energy shifts $E^{(1)}$
are summarized
in the last column of Table~\ref{tab_shift4} for the first three
energy manifolds with $E_{\mathrm{ni,rel}}=13 \hbar \omega/2$,
$15 \hbar \omega/2$ and
$17 \hbar \omega/2$.
Figure~\ref{fig_energyn4shiftbcs}
compares the four-fermion spectrum near
$E_{\mathrm{rel}}(2,2) \approx 13/2\hbar \omega$
calculated by the stochastic variational approach [squares, 
diamonds and pluses 
show the energy levels corresponding to states
with $(L_{\mathrm{rel}},\Pi_{\mathrm{rel}})=(0,+1)$, $(2,+1)$ and $(1,+1)$,
respectively]
with that calculated perturbatively
(solid and dashed lines show the energies
corresponding to states with natural and unnatural parity,
respectively).
As expected, the agreement is excellent
for small $|a^{(aa)}|$ and worsens with increasing $|a^{(aa)}|$.
For small $|a^{(aa)}|$, the energy level with
unnatural parity is affected less strongly by the two-body interactions
than the energy levels with natural parity.
Inspection of Table~\ref{tab_shift4}
shows that this is a general trend, i.e., within a given manifold
the energy level shifted most strongly is that
corresponding to the natural
parity state with the smallest allowed angular momentum. 

\begin{figure}
\vspace*{+.4cm}
\includegraphics[angle=0,width=70mm]{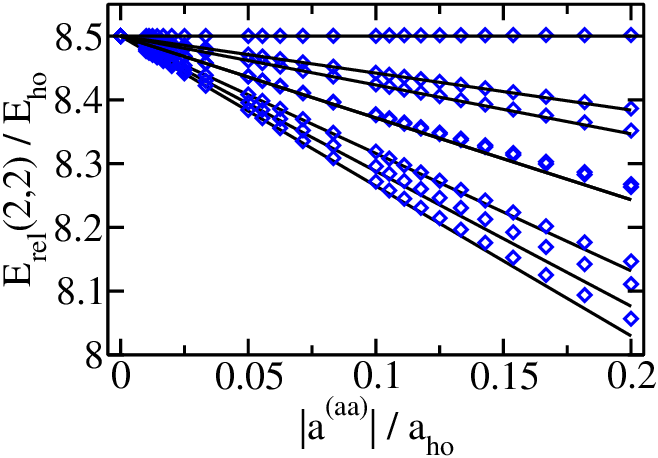}
\vspace*{0.1cm}
\caption{(Color online)
Four-fermion energies $E_{\mathrm{rel}}(2,2)$
as a function of the absolute value of the $s$-wave scattering length
$|a^{(aa)}|$ for small $|a^{(aa)}|$ ($a^{(aa)} \le 0$),
 $E_{\mathrm{rel}}(2,2) \approx 17 \hbar \omega/2$ and
$(L_{\mathrm{rel}},\Pi_{\mathrm{rel}})=(2,+1)$.
Diamonds
show the numerically determined four-fermion energies 
while solid 
lines
show the perturbative results.
Note that the fourth and fifth state (counted from
the bottom) are
nearly degenerate for the scattering length range depicted.
}\label{fig_energyn4shiftbcs2}
\end{figure}

To illustrate the behavior 
of the energy spectrum for a higher energy manifold,
Fig.~\ref{fig_energyn4shiftbcs2} shows 
the energies corresponding to 
the eight states with $(L_{\mathrm{rel}},\Pi_{\mathrm{rel}})=(2,+1)$
around $E_{\mathrm{rel}} \approx 17 \hbar \omega/2$.
Again, the agreement between the numerically determined energies (diamonds)
and the perturbatively determined
energies (solid lines) is excellent
for small $|a^{(aa)}|$.
Interestingly, 
within the perturbative treatment, one of the $(L_{\mathrm{rel}},\Pi_{\mathrm{rel}})=(2,+1)$
states is not affected by the 
$s$-wave interactions, implying that the wave function vanishes 
whenever two unlike fermions approach each other closely.
It turns out that the perturbative result in this case is exact, i.e.,
there exists a $(L_{\mathrm{rel}},\Pi_{\mathrm{rel}})=(2,+1)$
state with energy $E_{\mathrm{rel}}(2,2)=17 \hbar \omega/2$ for all
scattering lengths $a^{(aa)}$.

%Since the number of states increases dramatically as the energy increases,
%a density of states type approach suggests itself for visualizing the
%energy shifts of
%higher energy manifolds. 
Figures~\ref{fig_shiftn3}(c) and
(d) show the distributions of the $c^{(1)}$ coefficients
for the fourth and fifth energy manifolds with 
$E_{\mathrm{rel}}(2,2) \approx 19 \hbar \omega /2$ and
$E_{\mathrm{rel}}(2,2) \approx 21 \hbar \omega /2$, respectively.
Compared to the three-particle case
[Figs.~\ref{fig_shiftn3}(a) and (b)], the
degeneracies increase more rapidly as can be seen by the higher frequency
with which the $c^{(1)}$ coefficients occur.
Furthermore, the $c^{(1)}=0$ bin no longer dominates the distribution since
both natural and unnatural parity states are effected by the zero-range 
interactions.  

\begin{figure}
\vspace*{+.4cm}
\includegraphics[angle=0,width=70mm]{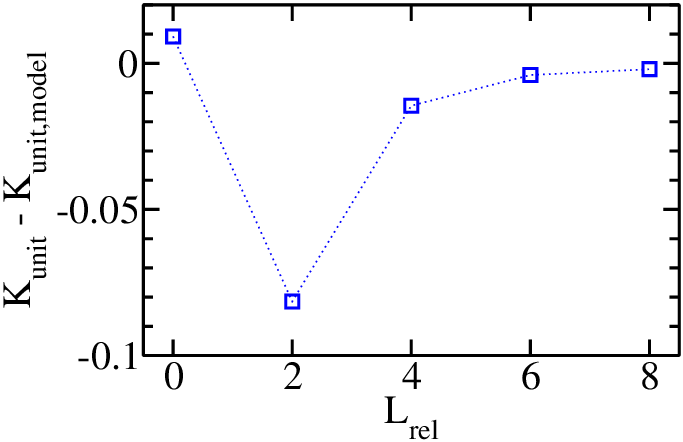}
\vspace*{0.1cm}
\caption{(Color online)
Symbols show
the difference between 
the lowest coefficient 
$K_{\mathrm{unit}}$
obtained from
our numerically determined four-fermion energies and 
the lowest coefficient 
$K_{\mathrm{unit,model}}$ predicted by
the simple dimer-dimer model
discussed in the text as a function of $L_{\mathrm{rel}}$ ($L_{\mathrm{rel}}$ even).
To guide the eye symbols are connected
by dotted lines.
}\label{fig_unit_coeff1}
\end{figure}

Next, we discuss the energy spectrum in the strongly-interacting regime and
in the $a^{(aa)} \rightarrow 0^+$ limit.
For small 
$a^{(aa)}$, the energetically lowest lying 
states with even $L_{\mathrm{rel}}$ belong to the ``dimer plus dimer''
family while the energetically lowest lying states with odd $L_{\mathrm{rel}}$
belong to the ``dimer plus atom plus atom'' family.
Motivated by this observation (see also 
Refs.~\cite{stec07c,stec08,stec08d})
and by our discussion of the three-fermion
system (see Sec.~\ref{sec_resultsn3}),
Figs.~\ref{fig_energyn4scale}(a), (b) and (c)
show the scaled energies $E_{\mathrm{rel}}(2,2)-2E_{\mathrm{dimer}}$
for $L_{\mathrm{rel}}=0,2$ and 4.
The 
scaled four-fermion spectra
for $L_{\mathrm{rel}}=2$ and 4 contain 
a set of nearly constant scaled energies
given by
$E_{\mathrm{rel}}(2,2)-2 E_{\mathrm{dimer}} \approx 
(3/2+L_{\mathrm{rel}}+2q)\hbar \omega$, where $q=0,1,\cdots$
[see dashed lines in Figs.~\ref{fig_energyn4scale}(b) and (c)].
Figures~\ref{fig_energyn4scale}(b) and (c) show that
this description improves with increasing $L_{\mathrm{rel}}$.
The set of constant scaled energies $E_{\mathrm{rel}}(2,2)-2E_{\mathrm{dimer}}$
is predicted to exist within the
effective 
two-boson model, which treats the
composite bosons as
non-interacting if $L_{\mathrm{rel}}>0$ and $L_{\mathrm{rel}}$ even [see the discussion
below
Eq.~(\ref{eq_busch}) in Sec.~\ref{sec_theory}]. 
Using that a subset of the scaled energies is approximately constant and
that the dimer energy $E_{\mathrm{dimer}}$ equals $\hbar \omega / 2$
at unitarity,
the lowest $K_{\mathrm{unit}}$
coefficient is, for $L_{\mathrm{rel}}$ even and $L_{\mathrm{rel}}>0$,
approximately
given by $K_{\mathrm{unit,model}}=L_{\mathrm{rel}}+1$.
The energy levels associated with these $K_{\mathrm{unit}}$ coefficients
have the minimally allowed number of excitations
in the hyperangular degrees of freedom 
and $q$ excitations along the hyperradial coordinate
$R$;
these states transform, as $a^{(aa)}$ changes from $\infty$
to $0^+$, to 
``dimer plus dimer'' states with $n_{\mathrm{eff}}$ radial excitations.
Since the dimers are composite 
bosons, an analogous set of states does
not exist for odd $L_{\mathrm{rel}}$.
Squares in Fig.~\ref{fig_unit_coeff1} 
show the difference between
the lowest $K_{\mathrm{unit}}$ coefficient and $K_{\mathrm{unit,model}}$, where $K_{\mathrm{unit}}$
is taken from Tables~\ref{tab_energyunit} and \ref{tab_energyunit2}
for $L_{\mathrm{rel}}=2-8$ and $L_{\mathrm{rel}}$ even.
This figure 
supports our conjecture that the lowest coefficient $K_{\mathrm{unit}}$ 
converges to
the value $K_{\mathrm{unit,model}}$ in the large
$L_{\mathrm{rel}}$ limit.

The behavior of the scaled energies for $L_{\mathrm{rel}}=0$ 
is different from that of the scaled energies
for $L_{\mathrm{rel}}>0$ ($L_{\mathrm{rel}}$ even).
Figure~\ref{fig_energyn4scale}(a) 
does not show a set of nearly constant scaled energy levels
but instead shows a set of scaled energies that drop by approximately
$ \hbar \omega$ in going from
$a^{(aa)} \rightarrow 0^-$ to $a^{(aa)} = \infty$
and by another
$ \hbar \omega$ in going from
$a^{(aa)} = \infty$ to $a^{(aa)} \rightarrow 0^+$.
Dashed lines 
in Fig.~\ref{fig_energyn4scale} show the energies predicted 
by the effective dimer-dimer model.
To obtain these energies, we 
solve Eq.~(\ref{eq_busch}) for $k=dd$
and use Eq.~(\ref{eq_dimerdimer})
to express the effective dimer-dimer 
scattering length $a^{(dd)}$
in terms of the atom-atom scattering length $a^{(aa)}$
throughout the entire crossover. The dashed lines agree
surprisingly well with the full four-fermion energies
throughout the entire crossover. Comparison with 
Fig.~\ref{fig_energyn3scaled}(a) shows that the effective two-particle
model provides a better description of the four-fermion than of the
three-fermion system, particularly for negative $a^{(aa)}$.
Intuitively, this might be explained by the fact that the 
three-fermion system contains an unpaired atom while all atoms
participate in the molecule formation in the four-fermion system. 
%More quantitative, ... NEED TO DO ...
Figure~\ref{fig_energyn4scale} 
suggests that the four-fermion energy spectrum
can be interpreted by considering avoided
crossings between dimer-dimer
states and states that belong to the
``dimer plus atom plus atom'' 
and the ``gas-like state'' families.
An analysis along these lines has been
performed in Refs.~\cite{stec07c,stec08d};
however, to the best of our knowledge this 
type of analysis has not been 
previously applied to 
estimate the value of the lowest $K_{\mathrm{unit}}$ coefficient
for $L_{\mathrm{rel}}=0$.
Using that the energy in the $a^{(aa)} \rightarrow 0^+$
limit is given by $2E_{\mathrm{dimer}}+(3/2 +2n_{\mathrm{eff}}) \hbar \omega$
and that the energy at unitarity is---using 
the argument above---$2E_{\mathrm{dimer}}+(5/2 +2q)\hbar \omega$,
the lowest $K_{\mathrm{unit}}$ coefficient 
for $L_{\mathrm{rel}}=0$ can be estimated to
be $K_{\mathrm{unit,model}}=2$.
Table~\ref{tab_energyunit} shows that this model is 
surprisingly accurate, i.e., $K_{\mathrm{unit}}=2.009$
(see also Fig.~\ref{fig_unit_coeff1}).

We find numerically that the second lowest 
$K_{\mathrm{unit}}$ coefficient
for even $L_{\mathrm{rel}}$ ($\Pi_{\mathrm{rel}}=+1$)
and the lowest 
$K_{\mathrm{unit}}$ coefficient for
odd $L_{\mathrm{rel}}$ ($\Pi_{\mathrm{rel}}=-1$) appear to approach 
$L_{\mathrm{rel}}+7/4$ for large $L_{\mathrm{rel}}$. At present,
we have no explanation for this observation.
While our analysis presented above allows for the prediction of 
the                   
lowest $K_{\mathrm{unit}}$ coefficient for all even 
$L_{\mathrm{rel}}$,       
%we have not been able to find similarly simple models that
%would explain the other $K_{unit}$ coefficients.
the remaining coefficients at unitarity appear to be result 
of an intricate interplay of how to best distribute the                  
angular momentum among the six interparticle distances.

\section{Summary}
\label{sec_conclusion}
This paper considered $s$-wave interacting
three- and four-fermion systems under external spherically
symmetric harmonic confinement.
If the range 
$r_0$ of the underlying interaction potential 
is much smaller than the other length scales of the system,
i.e., the atom-atom $s$-wave scattering length $a^{(aa)}$
and the harmonic oscillator length $a_{\mathrm{ho}}$,
then the 
physics of two-component equal-mass Fermi systems is universal.
%Experimentally,
%universal few-body 
%systems can be realized by loading cold atomic Fermi gases
%with two different hyperfine states
%into an optical lattice~\cite{grei02,koeh05,thal06} 
%or by loading a single dipole trap
%with just
%a few atoms~\cite{wilk10}, 
%and by adjusting 
%the trapping geometry and system parameters such
%that the van der Waals length (see., e.g., Ref.~\cite{wein99})
%and the effective range are much smaller than 
%$a^{(aa)}$ and $a_{\mathrm{ho}}$.
If the $s$-wave scattering length is tuned
in the vicinity of a so-called broad Feshbach 
resonance~\cite{dien04,petr05,gior08}, then
the low-energy properties of these few-fermion systems
are expected to behave universally, i.e.,
independently of the details of the underlying interaction
potential. Our results are expected to
apply to the experimentally
most frequently studied 
species $^6$Li and 
$^{40}$K~\cite{dema99,grei03,zwie03,bour03,stre03,kina04,stew06}. 

We have determined the zero-range energy spectrum of the 
three- and four-fermion systems numerically for a wide range 
of $s$-wave scattering lengths $a^{(aa)}$.
Our study of the three-fermion system is based on 
the Lippmann-Schwinger equation, which reduces the
problem
to solving a set
of coupled equations for each angular 
momentum $L_{\mathrm{rel}}$~\cite{kest07}.
The three-fermion spectrum is analyzed and interpreted
following a number of different approaches. In the weakly-interacting
regime, the interactions are treated as a perturbation
to the non-interacting trapped atomic Fermi gas.
This approach 
correctly 
describes
the gas-like states of the system 
for small $|a^{(aa)}|$,
with  $a^{(aa)}$ 
positive and negative,
but does not describe 
states that are associated with the formation of diatomic molecules.
The energies of this family of states are
described by an effective two-particle Hamiltonian that treats the
diatomic molecule as a composite point particle and assumes that the
spare atom and the molecule interact through the $s$-wave scattering 
channel characterized by the effective atom-dimer scattering length
$a^{(ad)}$.
Somewhat surprisingly, this effective two-particle model 
describes the energies of a subset of states
not only quantitatively correctly
in the weakly-repulsive regime but also qualitatively
for negative $a^{(aa)}$
(see dashed
lines in Fig.~\ref{fig_energyn3scaled}). Building on this
observation, we developed a simple model 
which predicts the $K_{\mathrm{unit}}$ coefficients
with reasonably good accuracy.
In doing so, our motivation was
to develop
a physical picture of the general features of
the three-fermion energy spectrum
that can, at
least partially, be generalized to larger systems. 
A key result of our analysis is that the energy levels
determined by the lowest 
$K_{\mathrm{unit}}$ coefficient for each $L_{\mathrm{rel}}$
transform to ``atom plus dimer'' states as $a^{(aa)}$ changes
from $\infty$ to $0^+$. This ``transformation''
of the states can be interpreted nicely
within the hyperspherical framework.
Using hyperspherical coordinates,
the states for $a^{(aa)} = \infty$ and $0^+$
are separable and the widths
of the avoided crossings of the energy levels
for finite $a^{(aa)}$ are determined by the couplings between
different hyperradial potential curves
(see, e.g., Refs.~\cite{delv58,delv60,mace68,lin95}
as well as Ref.~\cite{stec08d}).

We also solved the time-independent Schr\"odinger equation for the 
four-fermion system
for a Gaussian two-body potential with
varying range $r_0$
using the stochastic variational
approch. The resulting finite-range
energies were then extrapolated to the
$r_0 \rightarrow 0$ limit.
The energy spectrum of the four-fermion system is much denser
than that of the three-fermion system. Unlike for the three-fermion
system,
natural and unnatural parity states of the four-fermion
system are shifted by the zero-range
interactions.
Our primary focus in this paper has been to characterize the 
energies of the lowest few states with natural parity
throughout the crossover, including 
the infinitely strongly-interacting unitary regime.
As in the three-fermion case, our semi-analytical
perturbative approach, which utilizes hyperspherical coordinates,
reproduces the
numerically determined four-fermion 
energies with good accuracy in the weakly-attractive
and weakly-repulsive regimes. 
We paid special attention
to the infinitely strongly-interacting regime 
and analyzed the energy spectrum at unitarity within an 
effective two-boson model. This analysis 
provides a physical
picture of how the lowest $K_{\mathrm{unit}}$ coefficient for 
natural parity states and even $L_{\mathrm{rel}}$ comes about and
how the associated ladder of states transforms to ``dimer
plus dimer'' states as $a^{(aa)}\rightarrow 0^+$.
Furthermore, the fact that the effective two-boson model
reproduces a subset of energy levels semi-quantitatively 
throughout the
entire crossover regime suggests that the concept of the effective 
dimer-dimer scattering length $a^{(dd)}$
extends beyond the small $a^{(aa)}$ regime ($a^{(aa)}>0$).
%Hopefully??????????
%In addition, 
%we identify a family of $K_{\mathrm{unit}}$ coefficient that can 
%be interpreted using an effective 
%three-particle model that assumes the formation
%of one composite boson.

One motivation for studying small few-fermion systems is to
develop a ``bottom-up approach'' that investigates the microscopic physics
of two-component Fermi gases by treating successively larger systems.
With this motivation in mind,
the three-fermion system can be considered the smallest 
system that models spin-imbalanced Fermi gases~\cite{kest07} while
the four-fermion system can be considered the smallest 
system that models pairing physics 
throughout the BCS-BEC crossover
of spin-balanced Fermi gases~\cite{stec07c}.
In the future,
it will be interesting to extend the studies 
presented here to 
larger equal-mass 
few-fermion systems as well as to unequal-mass fermionic and
bosonic systems.
We also hope to use the perturbative treatment of the four-fermion
system to estimate the fourth order virial coefficient
in the weakly-interacting regimes.

Support by the NSF through
grant PHY-0855332
and by the ARO
are gratefully acknowledged.

%\bibliography{lit}       
%\bibliographystyle{prsty}

\newpage 

\begin{center}Auxiliary material for ``Energy spectrum of
harmonically trapped two-component Fermi gases:
Three- and Four-Particle Problem''
\end{center}

This EPAPS material contains the extrapolated zero-range 
energies $E_{\mathrm{rel}}(2,2)$ for negative scattering 
lengths $a^{(aa)}$,
and the extrapolated scaled 
zero-range energies $E_{\mathrm{rel}}(2,2)-2E_{\mathrm{dimer}}$ 
and
$E_{\mathrm{rel}}(2,2)-E_{\mathrm{dimer}}$ 
for positive scattering lengths $a^{(aa)}$. The notation follows that
of the main article.
\begin{table}
\caption{Energies $E_{\mathrm{rel}}(2,2)$ 
for various $L_{\mathrm{rel}}$ and $a^{(aa)}$.
The energies (second through 8th column) are reported in units
of
$E_{\mathrm{ho}}$.
Columns two through six show the energies for the five lowest 
states with $(L_{\mathrm{rel}},\Pi_{\mathrm{rel}})=(0,+1)$ while
columns seven through nine show the energies for the three lowest 
states with $(L_{\mathrm{rel}},\Pi_{\mathrm{rel}})=(1,-1)$.
The uncertainty of the energies
is estimated to be in the second to last digit for large
$|a^{(aa)}|$ and 1 in the last digit reported
for small $|a^{(aa)}|$.
``NI'' stands for non-interacting ($a^{(aa)}=0$).
}
\begin{ruledtabular}
\begin{tabular}{c|ccccc|ccc}
$a_{\mathrm{ho}}/a^{(aa)}$ & & &$L_{\mathrm{rel}}=0$ & & & &$L_{\mathrm{rel}}=1$ & \\ \hline 
   NI &   $13/2$ &   $17/2$ &   $17/2$ &   $17/2$ &   $17/2$ &   $15/2$ &   $15/2$ &   $15/2$ \\
$-100$ &   6.4741 &   8.4704 &   8.4721 &   8.4850 &   8.4885 &   7.4801 &   7.4824 &   7.4887 \\
 $-90$ &   6.4712 &   8.4671 &   8.4691 &   8.4834 &   8.4872 &   7.4779 &   7.4804 &   7.4875 \\
 $-80$ &   6.4676 &   8.4630 &   8.4652 &   8.4813 &   8.4856 &   7.4751 &   7.4780 &   7.4859 \\
 $-70$ &   6.4630 &   8.4577 &   8.4602 &   8.4787 &   8.4835 &   7.4716 &   7.4748 &   7.4840 \\
 $-60$ &   6.4569 &   8.4507 &   8.4536 &   8.4751 &   8.4808 &   7.4669 &   7.4707 &   7.4813 \\
 $-50$ &   6.4482 &   8.4409 &   8.4444 &   8.4701 &   8.4770 &   7.4603 &   7.4648 &   7.4775 \\
 $-40$ &   6.4354 &   8.4261 &   8.4306 &   8.4627 &   8.4712 &   7.4505 &   7.4561 &   7.4720 \\
 $-30$ &   6.4139 &   8.4016 &   8.4076 &   8.4503 &   8.4617 &   7.4342 &   7.4416 &   7.4627 \\
 $-20$ &   6.3714 &   8.3528 &   8.3620 &   8.4257 &   8.4427 &   7.4019 &   7.4127 &   7.4443 \\
 $-18$ &   6.3572 &   8.3366 &   8.3469 &   8.4176 &   8.4364 &   7.3913 &   7.4032 &   7.4382 \\
 $-16$ &   6.3397 &   8.3164 &   8.3281 &   8.4074 &   8.4285 &   7.3780 &   7.3913 &   7.4306 \\
 $-14$ &   6.3172 &   8.2906 &   8.3041 &   8.3944 &   8.4184 &   7.3611 &   7.3761 &   7.4208 \\
 $-12$ &   6.2873 &   8.2564 &   8.2723 &   8.3771 &   8.4050 &   7.3388 &   7.3560 &   7.4080 \\
 $-10$ &   6.2460 &   8.2090 &   8.2283 &   8.3532 &   8.3863 &   7.3080 &   7.3282 &   7.3901 \\
 $ -19/2$ &   6.2330 &   8.1931 &   8.2136 &   8.3455 &   8.3801 &   7.2983 &   7.3195 &   7.3846 \\
  $-9$ &   6.2187 &   8.1766 &   8.1984 &   8.3372 &   8.3736 &   7.2877 &   7.3099 &   7.3784 \\
  $-17/2$ &   6.2027 &   8.1583 &   8.1815 &   8.3280 &   8.3663 &   7.2759 &   7.2992 &   7.3715 \\
  $-8$ &   6.1848 &   8.1379 &   8.1626 &   8.3177 &   8.3581 &   7.2628 &   7.2873 &   7.3638 \\
  $-15/2$ &   6.1647 &   8.1149 &   8.1414 &   8.3060 &   8.3488 &   7.2480 &   7.2738 &   7.3552 \\
  $-7$ &   6.1419 &   8.0889 &   8.1174 &   8.2928 &   8.3382 &   7.2313 &   7.2586 &   7.3454 \\
  $-13/2$ &   6.1158 &   8.0592 &   8.0900 &   8.2778 &   8.3261 &   7.2123 &   7.2412 &   7.3342 \\
  $-6$ &   6.0857 &   8.0250 &   8.0584 &   8.2604 &   8.3119 &   7.1904 &   7.2213 &   7.3212 \\
  $-11/2$ &   6.0506 &   7.9853 &   8.0218 &   8.2401 &   8.2953 &   7.1650 &   7.1980 &   7.3062 \\
  $-5$ &   6.0091 &   7.9386 &   7.9788 &   8.2161 &   8.2755 &   7.1353 &   7.1706 &   7.2884 \\
  $-9/2$ &   5.9594 &   7.8830 &   7.9276 &   8.1875 &   8.2514 &   7.0998 &   7.1379 &   7.2671 \\
  $-4$ &   5.8989 &   7.8159 &   7.8658 &   8.1528 &   8.2216 &   7.0571 &   7.0983 &   7.2413 \\
  $-7/2$ &   5.8238 &   7.7337 &   7.7899 &   8.1097 &   8.1837 &   7.0046 &   7.0495 &   7.2092 \\
  $-3$ &   5.7284 &   7.6308 &   7.6948 &   8.0553 &   8.1342 &   6.9389 &   6.9879 &   7.1687 \\
  $-5/2$ &   5.6040 &   7.4994 &   7.5732 &   7.9845 &   8.0667 &   6.8544 &   6.9085 &   7.1158 \\
  $-2$ &   5.4361 &   7.3270 &   7.4133 &   7.8899 &   7.9700 &   6.7432 &   6.8027 &   7.0448 \\
  $-3/2$ &   5.2009 &   7.0936 &   7.1990 &   7.7588 &   7.8230 &   6.5916 &   6.6573 &   6.9456 \\
  $-1$ &   4.8569 &   6.7641 &   6.9062 &   7.5698 &   7.5847 &   6.3779 &   6.4497 &   6.8009 \\
  $-1/2$ &   4.3324 &   6.2745 &   6.5041 &   7.1829 &   7.2863 &   6.0663 &   6.1426 &   6.5796 \\
   $0$ &   3.5092 &   5.5101 &   5.9441 &   6.5290 &   6.8482 &   5.5978 &   5.6758 &   6.2305 \\
\end{tabular}
\end{ruledtabular}
\label{tab_bcs1}
\end{table}

\begin{table}
\caption{Nine lowest
energies $E_{\mathrm{rel}}(2,2)$ for $(L_{\mathrm{rel}},\Pi_{\mathrm{rel}})=(2,+1)$ and various $a^{(aa)}$.
The energies (second through 10th column) are reported in units
of
$E_{\mathrm{ho}}$.
The uncertainty of the energies
is estimated to be in the second to last digit for large
$|a^{(aa)}|$ and 1 in the last digit reported
for small $|a^{(aa)}|$.
}
\begin{ruledtabular}
\begin{tabular}{c|ccccccccc}
$a_{\mathrm{ho}}/a^{(aa)}$ & & & & &$L_{\mathrm{rel}}=2$ & & & &\\ \hline 
   NI &   13/2 &   17/2 &   17/2 &   17/2 &   17/2 &   17/2 &   17/2 &   17/2 &   17/2 \\
$-100$ &   6.4801 &   8.4766 &   8.4789 &   8.4816 &   8.4872 &   8.4872 &   8.4924 &   8.4942 &   8.5001 \\
$ -90$ &   6.4779 &   8.4740 &   8.4765 &   8.4796 &   8.4858 &   8.4858 &   8.4915 &   8.4936 &   8.5001 \\
$ -80$ &   6.4751 &   8.4706 &   8.4735 &   8.4769 &   8.4838 &   8.4838 &   8.4903 &   8.4927 &   8.5001 \\
$ -70$ &   6.4716 &   8.4665 &   8.4698 &   8.4737 &   8.4817 &   8.4817 &   8.4891 &   8.4917 &   8.5001 \\
$ -60$ &   6.4669 &   8.4610 &   8.4648 &   8.4694 &   8.4787 &   8.4787 &   8.4873 &   8.4903 &   8.5001 \\
$ -50$ &   6.4604 &   8.4532 &   8.4579 &   8.4633 &   8.4745 &   8.4745 &   8.4847 &   8.4884 &   8.5001 \\
$ -40$ &   6.4506 &   8.4416 &   8.4474 &   8.4541 &   8.4681 &   8.4682 &   8.4809 &   8.4855 &   8.5002 \\
$ -30$ &   6.4344 &   8.4223 &   8.4300 &   8.4388 &   8.4576 &   8.4578 &   8.4746 &   8.4807 &   8.5002 \\
$ -20$ &   6.4023 &   8.3841 &   8.3957 &   8.4085 &   8.4367 &   8.4372 &   8.4619 &   8.4711 &   8.5003 \\
$ -18$ &   6.3918 &   8.3714 &   8.3843 &   8.3984 &   8.4298 &   8.4304 &   8.4577 &   8.4679 &   8.5003 \\
$ -16$ &   6.3786 &   8.3556 &   8.3703 &   8.3858 &   8.4213 &   8.4220 &   8.4524 &   8.4639 &   8.5004 \\
$ -14$ &   6.3619 &   8.3354 &   8.3522 &   8.3697 &   8.4104 &   8.4112 &   8.4457 &   8.4588 &   8.5004 \\
$ -12$ &   6.3398 &   8.3087 &   8.3285 &   8.3484 &   8.3959 &   8.3970 &   8.4367 &   8.4520 &   8.5005 \\
$ -10$ &   6.3094 &   8.2717 &   8.2956 &   8.3187 &   8.3760 &   8.3774 &   8.4243 &   8.4425 &   8.5005 \\
$  -19/2$ &   6.2992 &   8.2579 &   8.2841 &   8.3079 &   8.3695 &   8.3712 &   8.4205 &   8.4394 &   8.5012 \\
$  -9$ &   6.2887 &   8.2450 &   8.2728 &   8.2976 &   8.3626 &   8.3645 &   8.4161 &   8.4361 &   8.5013 \\
$  -17/2$ &   6.2771 &   8.2306 &   8.2602 &   8.2861 &   8.3550 &   8.3570 &   8.4113 &   8.4324 &   8.5013 \\
$  -8$ &   6.2642 &   8.2145 &   8.2462 &   8.2732 &   8.3464 &   8.3487 &   8.4058 &   8.4283 &   8.5014 \\
$  -15/2$ &   6.2494 &   8.1964 &   8.2304 &   8.2587 &   8.3368 &   8.3393 &   8.3997 &   8.4235 &   8.5014 \\
$  -7$ &   6.2329 &   8.1759 &   8.2127 &   8.2422 &   8.3260 &   8.3287 &   8.3927 &   8.4182 &   8.5015 \\
$  -13/2$ &   6.2141 &   8.1525 &   8.1924 &   8.2234 &   8.3136 &   8.3166 &   8.3848 &   8.4120 &   8.5016 \\
$  -6$ &   6.1925 &   8.1254 &   8.1692 &   8.2018 &   8.2994 &   8.3028 &   8.3755 &   8.4049 &   8.5016 \\
$  -11/2$ &   6.1675 &   8.0939 &   8.1423 &   8.1765 &   8.2828 &   8.2867 &   8.3646 &   8.3965 &   8.5017 \\
$  -5$ &   6.1380 &   8.0566 &   8.1109 &   8.1467 &   8.2634 &   8.2678 &   8.3518 &   8.3864 &   8.5018 \\
$  -9/2$ &   6.1030 &   8.0121 &   8.0735 &   8.1112 &   8.2402 &   8.2453 &   8.3362 &   8.3743 &   8.5020 \\
$  -4$ &   6.0607 &   7.9581 &   8.0287 &   8.0680 &   8.2122 &   8.2181 &   8.3172 &   8.3593 &   8.5021 \\
$  -7/2$ &   6.0087 &   7.8911 &   7.9738 &   8.0147 &   8.1777 &   8.1847 &   8.2933 &   8.3402 &   8.5022 \\
$  -3$ &   5.9432 &   7.8064 &   7.9053 &   7.9475 &   8.1343 &   8.1428 &   8.2627 &   8.3155 &   8.5024 \\
$  -5/2$ &   5.8587 &   7.6964 &   7.8178 &   7.8606 &   8.0782 &   8.0888 &   8.2219 &   8.2819 &   8.5025 \\
$  -2$ &   5.7459 &   7.5496 &   7.7027 &   7.7451 &   8.0035 &   8.0170 &   8.1658 &   8.2342 &   8.5028 \\
$  -3/2$ &   5.5895 &   7.3473 &   7.5456 &   7.5865 &   7.9001 &   7.9180 &   8.0847 &   8.1620 &   8.5030 \\
$  -1$ &   5.3613 &   7.0597 &   7.3211 &   7.3615 &   7.7501 &   7.7757 &   7.9610 &   8.0430 &   8.5033 \\
$  -1/2$ &   5.0077 &   6.6422 &   6.9799 &   7.0320 &   7.5194 &   7.5610 &   7.7613 &   7.8277 &   8.5036 \\
$   0$ &   4.4185 &   6.0390 &   6.4201 &   6.5391 &   7.1288 &   7.2218 &   7.4247 &   7.4271 &   8.0416 \\
\end{tabular}
\end{ruledtabular}
\label{tab_bcs2}
\end{table}

\begin{table}
\caption{Energies $E_{\mathrm{rel}}(2,2)$ for various $L_{\mathrm{rel}}$ and $a^{(aa)}$.
The energies (second through 5th column) are reported in units
of
$E_{\mathrm{ho}}$.
Column two shows the energy for the lowest 
state with $(L_{\mathrm{rel}},\Pi_{\mathrm{rel}})=(3,-1)$ while
columns three through five show the energies for the three lowest 
states with $(L_{\mathrm{rel}},\Pi_{\mathrm{rel}})=(4,+1)$.
The uncertainty of the energies
is estimated to be in the second to last digit for large
$|a^{(aa)}|$ and 1 in the last digit reported
for small $|a^{(aa)}|$.}
\begin{ruledtabular}
\begin{tabular}{c|c|ccc}
$a_{\mathrm{ho}}/a^{(aa)}$ & $L_{\mathrm{rel}}=3$ &  &$L_{\mathrm{rel}}=4$ & 
\\ \hline
NI &   15/2 &   17/2 &   17/2 &   17/2 \\
$-100$ &   7.4861 &   8.4823 &   8.4901 &   8.4908 \\
$ -90$ &   7.4845 &   8.4804 &   8.4890 &   8.4898 \\
$ -80$ &   7.4826 &   8.4778 &   8.4874 &   8.4884 \\
$ -70$ &   7.4802 &   8.4748 &   8.4858 &   8.4869 \\
$ -60$ &   7.4769 &   8.4706 &   8.4835 &   8.4847 \\
$ -50$ &   7.4723 &   8.4648 &   8.4802 &   8.4817 \\
$ -40$ &   7.4655 &   8.4561 &   8.4753 &   8.4771 \\
$ -30$ &   7.4541 &   8.4416 &   8.4671 &   8.4695 \\
$ -20$ &   7.4318 &   8.4130 &   8.4509 &   8.4544 \\
$ -18$ &   7.4244 &   8.4036 &   8.4456 &   8.4494 \\
$ -16$ &   7.4153 &   8.3918 &   8.4389 &   8.4432 \\
$ -14$ &   7.4036 &   8.3769 &   8.4304 &   8.4352 \\
$ -12$ &   7.3882 &   8.3571 &   8.4192 &   8.4247 \\
$ -10$ &   7.3671 &   8.3298 &   8.4036 &   8.4100 \\
$  -19/2$ &   7.3602 &   8.3206 &   8.3987 &   8.4053 \\
$  -9$ &   7.3529 &   8.3112 &   8.3933 &   8.4002 \\
$  -17/2$ &   7.3449 &   8.3007 &   8.3873 &   8.3945 \\
$  -8$ &   7.3359 &   8.2890 &   8.3807 &   8.3881 \\
$  -15/2$ &   7.3258 &   8.2759 &   8.3732 &   8.3810 \\
$  -7$ &   7.3144 &   8.2610 &   8.3646 &   8.3728 \\
$  -13/2$ &   7.3014 &   8.2440 &   8.3549 &   8.3636 \\
$  -6$ &   7.2865 &   8.2245 &   8.3437 &   8.3529 \\
$  -11/2$ &   7.2693 &   8.2018 &   8.3307 &   8.3403 \\
$  -5$ &   7.2490 &   8.1751 &   8.3153 &   8.3255 \\
$  -9/2$ &   7.2250 &   8.1432 &   8.2969 &   8.3077 \\
$  -4$ &   7.1961 &   8.1046 &   8.2745 &   8.2860 \\
$  -7/2$ &   7.1606 &   8.0569 &   8.2469 &   8.2588 \\
$  -3$ &   7.1162 &   7.9965 &   8.2119 &   8.2242 \\
$  -5/2$ &   7.0593 &   7.9179 &   8.1662 &   8.1786 \\
 $ -2$ &   6.9841 &   7.8121 &   8.1046 &   8.1163 \\
 $ -3/2$ &   6.8812 &   7.6630 &   8.0182 &   8.0275 \\
 $ -1$ &   6.7346 &   7.4409 &   7.8909 &   7.8944 \\
 $ -1/2$ &   6.5164 &   7.0871 &   7.6837 &   7.6926 \\
  $ 0$ &   6.1756 &   6.4855 &   7.3384 &   7.3677 \\
\end{tabular}
\end{ruledtabular}
\label{tab_bcs3}
\end{table}

\begin{table}
\caption{
Scaled energies $E_{\mathrm{rel}}(2,2)-2E_{\mathrm{dimer}}$ for various 
$L_{\mathrm{rel}}$ and $a^{(aa)}$.
The scaled energies (second to 10th column) are reported in units
of
$E_{\mathrm{ho}}$.
Columns two through four show the 
scaled energies for the three lowest 
states with $(L_{\mathrm{rel}},\Pi_{\mathrm{rel}})=(0,+1)$,
columns five through seven show the 
scaled energies for the three lowest 
states with $(L_{\mathrm{rel}},\Pi_{\mathrm{rel}})=(2,+1)$
and
columns eight through ten show the 
scaled energies for the three lowest 
states with $(L_{\mathrm{rel}},\Pi_{\mathrm{rel}})=(4,+1)$.
The uncertainty of the scaled energies
is estimated to be in the last digit reported.}
\begin{ruledtabular}
\begin{tabular}{c|ccc|ccc|ccc}
$a_{\mathrm{ho}}/a^{(aa)}$ & & $L_{\mathrm{rel}}=0$ &&&$L_{\mathrm{rel}}=2$ &&&$L_{\mathrm{rel}}=4$ &\\ \hline 
$   1/2$ & 2.281 & 4.357 & 5.190 & 3.457 & 5.239 & 5.539 & 5.495 & 6.836 & 6.890 \\
$   1$ & 2.082 & 4.206 & 5.718 & 3.486 & 5.464 & 6.073 & 5.499 & 7.493 & 7.719 \\
$   3/2$ & 1.939 & 4.072 & 6.138 & 3.497 & 5.486 & 7.264 & 5.500 & 7.503 & 9.033 \\
$   2$ & 1.843 & 3.967 & 6.045 & 3.499 & 5.496 & 7.500 & 5.500 & 7.501 & 9.508 \\
$   5/2$ & 1.778 & 3.888 & 5.963 & 3.501 & 5.499 & 7.504 & 5.500 & 7.501 & 9.506 \\
$   3$ & 1.732 & 3.831 & 5.899 & 3.500 & 5.500 & 7.506 & 5.500 & 7.501 & 9.504 \\
$   7/2$ & 1.699 & 3.787 & 5.848 & 3.501 & 5.500 & 7.503 & 5.500 & 7.501 & 9.504 \\
$   4$ & 1.673 & 3.752 & 5.807 & 3.502 & 5.500 & 7.502 & 5.500 & 7.501 & 9.501 \\
$   9/2$ & 1.653 & 3.724 & 5.773 & 3.501 & 5.500 & 7.502 & 5.500 & 7.501 & 9.501 \\
$   5$ & 1.637 & 3.701 & 5.745 & 3.500 & 5.500 & 7.502 & 5.500 & 7.501 & 9.501 
\end{tabular}
\end{ruledtabular}
\label{tab_bec1}
\end{table}

\begin{table}
\caption{Scaled energies $E_{\mathrm{rel}}(2,2)-E_{\mathrm{dimer}}$ 
for various 
$L_{\mathrm{rel}}$ and $a^{(aa)}$.
The scaled energies (second through 5th column) are reported in units
of
$E_{\mathrm{ho}}$.
Columns two through four show the 
scaled energies for the three lowest 
states with $(L_{\mathrm{rel}},\Pi_{\mathrm{rel}})=(1,-1)$
and
column five shows the scaled energy for the lowest 
state with $(L_{\mathrm{rel}},\Pi_{\mathrm{rel}})=(3,-1)$.
The uncertainty of the scaled energies
is estimated to be in the last digit reported.}
\begin{ruledtabular}
\begin{tabular}{c|ccc|c}
$a_{\mathrm{ho}}/a^{(aa)}$ & & $L_{\mathrm{rel}}=1$ & &$L_{\mathrm{rel}}=3$ \\ \hline 
$   1/2$ & 4.927 & 4.999 & 5.703 & 5.667 \\
$   1$ & 4.774 & 4.834 & 5.708 & 5.697 \\
$   3/2$ & 4.654 & 4.699 & 5.745 & 5.753 \\
$   2$ & 4.562 & 4.593 & 5.795 & 5.813 \\
$   5/2$ & 4.490 & 4.511 & 5.841 & 5.863 \\
$   3$ & 4.433 & 4.446 & 5.880 & 5.900 \\
$   7/2$ & 4.387 & 4.395 & 5.909 & 5.926 \\
$   4$ & 4.350 & 4.352 & 5.931 & 5.945 \\
$   9/2$ & 4.318 & 4.318 & 5.947 & 5.958 \\
$   5$ & 4.289 & 4.290 & 5.959 & 5.968 
\end{tabular}
\end{ruledtabular}
\label{tab_bec2}
\end{table}

\end{document}